\documentclass[twocolumn,tightenlines,aps,preprintnumbers,superscriptaddress,showpacs,floatfix]{revtex4} 
\usepackage{graphicx}
\usepackage{epsf}
\usepackage{amssymb}

\newcommand{\lgtr}{\scriptscriptstyle\lessgtr}

\begin{document}
\preprint{LA-UR-04-8611}
\preprint{DO-TH-04/14}
\title{Inflationary Perturbations and Precision Cosmology}
\author{Salman Habib}
\affiliation{T-8, The University of California,
Los Alamos National Laboratory, Los Alamos, 
New Mexico 87545, USA}

\author{Andreas Heinen}
\affiliation{Institut f\"ur Physik, Universit\"at Dortmund,
D-44221 Dortmund, Germany}

\author{Katrin Heitmann}
\affiliation{ISR-1, The University of California,
Los Alamos National Laboratory, Los Alamos, 
New Mexico 87545, USA}

\author{Gerard Jungman} 
\affiliation{T-6, The University of California,
Los Alamos National Laboratory, Los Alamos, New Mexico 87545, USA}

\date{\today}

\begin{abstract}
  
  Inflationary cosmology provides a natural mechanism for the
  generation of primordial perturbations which seed the formation of
  observed cosmic structure and lead to specific signals of anisotropy
  in the cosmic microwave background radiation.  In order to test the
  broad inflationary paradigm as well as particular models against
  precision observations, it is crucial to be able to make accurate
  predictions for the power spectrum of both scalar and tensor
  fluctuations.  We present detailed calculations of these quantities
  utilizing direct numerical approaches as well as error-controlled
  uniform approximations, comparing with the (uncontrolled)
  traditional slow-roll approach.  A simple extension of the
  leading-order uniform approximation yields results for the power
  spectra amplitudes, the spectral indices, and the running of
  spectral indices, with accuracy of the order of $0.1\%$ --
  approximately the same level at which the transfer functions are
  known.  Several representative examples are used to demonstrate
  these results.

\end{abstract}


\maketitle

\section{Introduction}

Cosmological inflation~\cite{infrefs} is a central component of the
present theoretical picture of cosmology. Inflation not only directly
addresses and solves fundamental weaknesses of the older Big Bang
picture -- the flatness and horizon problems -- it also provides an
elegant mechanism for the creation of primordial
fluctuations~\cite{infpert}, essential for explaining the observed
structure of the present-day Universe. As the Universe inflates,
microscopic quantum scales are stretched to macroscopic cosmological
scales. Quantum vacuum fluctuations provide the initial seeds that are
amplified by gravitational instability, leading to the formation of
structure in the Universe.

The primordial fluctuations associated with inflation are, primarily,
of a very simple type. In ``standard'' inflationary models, they arise
from the fluctuations of an effectively free scalar field and are
hence Gaussian random fields, completely characterized by two-point
statistics, such as the power spectrum. The basic task, then, is to
determine the scalar and tensor perturbations (vector perturbations
being naturally suppressed) in terms of the associated power spectra.
Power spectra from standard inflation models are conveniently
parameterized by a spectral index and its (weak) variation with scale
-- the ``running'' of the spectral index. In addition, inflation
couples the generation of scalar and tensor fluctuations; if both are
independently measured, ``compatibility'' relations characteristic of
inflation can be put to observational test. It should be noted that
complicated inflationary models can be constructed to produce baroque
``designer'' power spectra. Present observations do not require
building such models.

\begin{figure}[htbp]
  \centering
  \includegraphics[width=1.00\columnwidth]{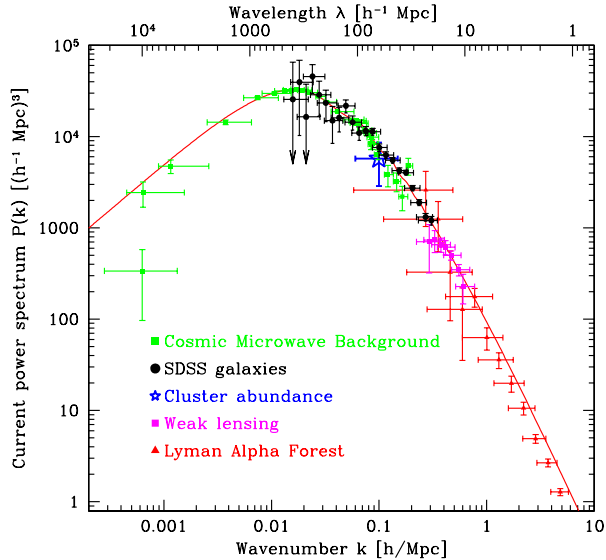}
  \caption{Compilation of observations of the matter power spectrum
  $P_{\rm obs}(k)$ taken from Tegmark et al.~\cite{tegmark} (with
  permission).  This particular figure assumes a flat scalar
  scale-invariant model with $\Omega_m=0.28$, $h=0.72$ and
  $\Omega_b/\Omega_m=0.16$, $\tau=0.17$, and a bias, $b_*=0.92$ for
  the Sloan Digital Sky Survey (SDSS) galaxies. The solid line is the
  theoretical curve.}
  \label{obsfig}
\end{figure}

Observational constraints on the primordial power spectra arise from
measurements of temperature anisotropies in the cosmic microwave
background radiation (CMBR), sensitive to both scalar and tensor
modes, and from measurements of the density power spectrum in surveys
of the large-scale mass distribution in the Universe. Ground and
satellite-based observations of the CMBR anisotropy have yielded
results very consistent with the essential adiabatic and Gaussian
nature of inflationary perturbations, with a value of the scalar
spectral index very close to unity~\cite{cmbobs}. The tensor
fluctuations are expected to be much lower in amplitude and have not
yet been observed. Scalar perturbations seed structure formation and
hence can be measured by observing the large-scale distribution of
galaxies and neutral hydrogen, as in ongoing redshift
surveys~\cite{lssobs} and Ly$\alpha$ observations~\cite{lyalpha}. This
second set of independent measurements provides information on smaller
scales than the CMBR, yet there is an overlap region where both
measurements have been shown to be consistent (Fig.~\ref{obsfig}). The
observed power spectrum is $P_{\rm obs}(k)=P(k)T^2(k)$ where $P(k)$ is
the primordial power spectrum and $T(k)$ is the transfer function for
radiation or matter as appropriate.

As measurements continue to improve, tests of the inflationary
paradigm and specific inflationary models will become more stringent,
especially if accurate measurements of the running of the spectral
index are performed and tensor perturbations are observed.  In
addition, an inverse analysis of the observational data may then be
attempted, in an effort to directly measure the inflationary
``equation of state''~\cite{hhj}.  In keeping with the remarkable
improvement of observational capabilities, the quality of theoretical
predictions has necessarily to be addressed.  The central theme of
this paper is to quantify and improve the quality of theoretical
predictions.

Scalar and tensor power spectra for particular inflation models can be
calculated either by direct numerical methods or by employing analytic
or semi-analytic approximations such as the slow-roll
approach~\cite{slowrev} or uniform approximations~\cite{hhjm, hhhjm}. 
For individual models, the numerical approach may well be preferable,
but powerful approximations have their own advantages.  They provide
intuition and understanding applicable to entire classes of models. 
Provided tight error controls can be met, they are much faster than a
mode-by-mode numerical integration for obtaining the power spectrum. 
Also, both numerical and approximate strategies can be melded together
by first obtaining a set of approximate results over a wide range of
parameters and then spot-checking numerically.  It is our view that
both strategies should be applied and compared against each other as
they have differing error modes.  Finally, it should be kept in mind
that the ultimate accuracy with which the primordial fluctuations must
be computed is limited by how accurately the radiation and matter
transfer functions are known, as well as by the errors associated with
observations.  At present, the transfer functions can be computed to
about $0.1\%$ accuracy~\cite{sswz}, sufficient for dealing with
next-generation observations.  We will use this value as a relative
figure of merit when discussing errors below.  It should be noted that
the accuracy requirements for the tensor or gravitational wave
component are not as stringent as for the scalar component; the
expected signal is at large length scales, where cosmic variance
becomes unavoidable.  Finally, measurements of the radiation and
matter power spectrum at higher values of $k$ face issues such as the
rapid fall-off of the primary signal, secondary point-source
contamination, foreground subtraction (for the CMBR), galaxy bias,
systematic errors, effects from baryons~\cite{baryons} and neutrinos
\cite{aba}, and limited accuracy in theoretical computation of the
present nonlinear power spectrum~\cite{codecomp} for the matter
distribution.

In this paper we focus on the perturbation spectrum for single-field
inflationary models.  We implement a direct numerical approach as well
as the slow-roll and uniform approximations, with the aim of
establishing control over the errors associated with each method and
understanding their associated advantages and disadvantages.  The main
technical advances are a robust and numerically efficient strategy for
the mode-by-mode integration, simple and useful error estimates for
the uniform approximation, and a simple improvement strategy for power
spectra amplitudes for the uniform approximation at leading order.  We
are able to show that the improved leading order uniform approximation
leads to very good accuracy for the spectral indices and their
running, as well as for the amplitudes of the power spectra and the
ratio of tensor to scalar perturbations.  For the most part, the
accuracy of the results obtained is of the order of $0.1\%$.

The method of uniform approximation employed in this paper has been
presented in Refs.~\cite{hhjm, hhhjm}. This method is a
``uniformization'' of the well-known Wentzel-Kramers-Brillouin (WKB)
or Liouville-Green (LG) approximation~\cite{lg} in the presence of
transition points. The uniform approximation began with the work of
Langer~\cite{langer} and others, and was followed by the notable
contributions of Olver~\cite{olverpap1,Olver} which we rely on for our
analysis below. While this line of investigation rested on the
analysis of ordinary differential equations, equivalent asymptotic
results based on an integral representation were given by Chester,
Friedman, and Ursell~\cite{uinteg}. The uniform approximation has
proven very useful in many applications, e.g., chemical
physics~\cite{cm}, the semiclassical limit in quantum
mechanics~\cite{berry}, and the quantum-classical transition in
quantum cosmology~\cite{shua}. For our purposes, the key advantages of
the Olver uniform approximation are that it does away with WKB-like
matching conditions~\cite{jeff} (such a procedure fails to indicate
the error of the approximation), has controlled error bounds over the
entire domain of interest, is systematically improvable, and allows
analytic simplifications in special cases of physical interest.

The paper is organized as follows. Section \ref{background} provides
the necessary background regarding the calculation of the primordial
power spectra and spectral indices for single field inflation models.
In Section \ref{uniapproximation} the results from the uniform
approximation in leading order are given, while in Section
\ref{slowroll} the essential equations for the slow-roll approximation
are summarized.  In Section \ref{numerics} we give a detailed
description of the numerical determination of the primordial power
spectra and spectral indices by solving the mode-equations
numerically, in the uniform approximation, and in the slow-roll
approximation. We investigate three different examples in Section
\ref{examples} and conclude with a discussion of our results in
Section \ref{conclusion}.

\section{Background}
\label{background}
The generation of perturbations during inflation is due to the
amplification of quantum vacuum fluctuations by the dynamics of the
background spacetime.  In order to calculate the perturbation spectra
for these quantum vacuum fluctuations three steps are necessary: the
dynamics of the background spacetime must be determined, the mode
equations for the scalar and tensor perturbations must be solved,
and finally, the power spectra themselves must be calculated as
functions of wave-number $k$. We will briefly outline these three
steps in the following; for detailed derivations the reader is refered
to the literature~\cite{slowrev,ivrevs2,LLKCBA,dodelson}.

\subsection{Background Equations}
For single-field inflationary models, i.e., when inflation is driven 
by a single homogeneous scalar field, the dynamical equation for the
inflaton field $\phi$ is given by 
\begin{equation}\label{eomphi}
\ddot{\phi}(t)+3H(t)\dot{\phi}(t)+\frac{\partial V(\phi)}{\partial\phi}=0.
\end{equation}
The dots denote derivatives with respect to physical time~$t$.  The
evolution of the Hubble parameter $H(t)=\dot a(t)/a(t)$, where $a(t)$ 
is the scale factor, is given by the Friedmann equation
\begin{equation}\label{hsq}
H^2(t)=\frac{8\pi G}{3}\left[\frac{1}{2}\dot{\phi}^2(t)+V(\phi)\right].
\end{equation}
Equivalently, by taking the derivative of $H(t)$ with respect to time
and inserting the equation of motion (\ref{eomphi}) for $\phi$ we can
write
\begin{equation}\label{hdot}
\dot H(t)=-4\pi G \dot\phi^2(t).
\end{equation}
Here and in the following we have set $\hbar=c=1$. 
Eqns.~(\ref{eomphi}) and (\ref{hsq}) or (\ref{hdot}) determine the
evolution of the inflaton field completely.  In addition, it is
useful to calculate the number of expansion e-folds via
\begin{equation}
\dot{N}(t)=H(t).
\end{equation}
Monitoring the number of e-folds for which inflation lasts allows us
to select observationally relevant inflationary models and their
parameters.  Finally, the conformal time $\eta(t)$ is defined by
\begin{equation}
\dot{\eta}(t)=\frac{1}{a(t)}.
\end{equation}
As explained later, the uniform approximation is expressed naturally
in terms of the conformal time.

Once the background equations are solved we can proceed to the next
step, the evaluation of the scalar and tensor perturbations.

\subsection{Mode Equations, Power Spectra, Spectral Indices, and their 
  Running} The modern understanding of fluctuations generated by
inflation is based on the gauge-invariant treatment of linearized
fluctuations in the metric and field
quantities~\cite{ivrevs2,givb,givs,ivrevs1}.  A particularly
convenient quantity for characterizing the perturbations is the
intrinsic curvature perturbation of the comoving
hypersurfaces~\cite{givdhl}, $\zeta\equiv u/z$, where $u$ is a
gauge-invariant scalar perturbation~\cite{ivrevs2}, and
$z\equiv a/(c_s H)[-\dot H/(4\pi G)+{\mathcal K}/a]^{1/2}$, where
$c_s$ is the sound speed and ${\mathcal K}$ is the curvature of
spatial sections. For single-field inflationary models, this
simplifies to $z= a\dot\phi/H$. The quantity $u$ satisfies the
dynamical equation 
\begin{equation}
u''-\Delta u - {z''\over z}u=0,
\label{uevol}
\end{equation}
where primes denote derivatives with respect to conformal time and 
$\Delta$ is the spatial Laplacian in comoving coordinates. It
follows immediately that $\zeta$ is approximately constant in the long
wavelength limit $k\rightarrow 0$.  This is true during the
inflationary phase as well as during the post-reheating era.  Moreover,
the Einstein equations can be used to connect the gravitational
potential $\Phi_A$ and $\zeta$ so that a computation of the power
spectrum of $\zeta$ provides all the information needed (aside from
the transfer functions) to extract the temperature anisotropy of the
CMBR. Details of this procedure can be found in
Refs.~\cite{ivrevs2,ms}.

The calculation of the relevant power spectra involves a computation
of the two-point functions for the appropriate quantum operators,
e.g., 
\begin{equation}
\langle 0|\hat{u}(\eta,{\bf x})\hat{u}(\eta,{\bf x}+{\bf
r})|0\rangle=\int_0^{\infty} {dk\over k} {\sin kr\over kr} P_u(\eta,k),
\label{pspectra}
\end{equation}
the operator $\hat{u}$ being written as
\begin{equation}
\hat{u}(x) = \int {d^3k\over (2\pi)^{3/2}}\left[\hat{a}_k u_k(\eta)
\hbox{e}^{i{\bf k}\cdot{\bf x}} + \hat{a}_k^{\dagger}
u_k^*(\eta)\hbox{e}^{-i{\bf k}\cdot{\bf x}}\right] ,
\label{uexpand}
\end{equation}
where $\hat{a}_k,~\hat{a}_k^{\dagger}$ are annihilation and creation
operators with $[\hat{a}_k,\hat{a}_{k'}^{\dagger}]
= \delta_{kk'}$, and $\hat{a}_k|0\rangle = 0$ $\forall k$.
The complex amplitude $u_k(\eta)$ satisfies
\begin{equation}
u_k^{\prime\prime}+\left(k^2 -{z^{\prime\prime}\over z}\right)u_k=0. 
\label{modeu}
\end{equation}
Solving Eqn.~(\ref{modeu}) is the fundamental problem in determining
the primordial power spectrum. The
corresponding mode equation for tensor perturbations is given by
\begin{equation}
v_k^{\prime\prime}+\left(k^2 -{a^{\prime\prime}\over a}\right)v_k=0. 
\label{modev}
\end{equation}

Once the mode equations are solved for different momenta $k$
the power spectra for the scalar and tensor modes are obtained via
\begin{eqnarray}\label{PS}
P_S(k)=\lim_{k\eta\rightarrow 
0^-}\frac{k^3}{2\pi^2}\left|\frac{u_k(\eta)}{z(\eta)}\right|^2,\\    
\label{PT}
P_T(k)=\lim_{k\eta\rightarrow 
0^-}\frac{k^3}{2\pi^2}\left|\frac{v_k(\eta)}{a(\eta)}\right|^2,   
\end{eqnarray}
where we denote the power spectrum for $\zeta$ by $P_S$.  

The tensor power spectrum is often defined with an additional factor
as
\begin{equation} 
P_h(k)=8P_T(k),
\end{equation}
leading to the definition of the tensor to scalar ratio
as
\begin{equation}\label{ratio}
R(k)=\frac{P_h(k)}{P_S(k)}=\frac{8P_T(k)}{P_S(k)}.
\end{equation}

The generalized spectral indices for the scalar and tensor perturbations 
are defined to be
\begin{eqnarray}
\label{defns}    
n_S(k)&=&1+\frac{d\ln P_S(k)}{d\ln k},   \\
\label{defnt}    
n_T(k)&=&\frac{d\ln P_T(k)}{d\ln k}.    
\end{eqnarray}    
Running of the spectral indices is conventionally parameterized by 
the second logarithmic derivative of the power spectra:
\begin{eqnarray}
\alpha_S(k)&=&\frac{d\ln n_S(k)}{d\ln k},\\
\alpha_T(k)&=&\frac{d\ln n_T(k)}{d\ln k}.
\end{eqnarray}
Numerical results for the power spectrum are often fit in the
literature assuming a power-law behavior and a small running of the
spectral index around a pivot scale $k_*$ (see e.g.,
Refs.~\cite{LLKCBA,LLMS}).  This fitting ansatz for the power spectrum
is
\begin{equation}
P(k)=A_\mathrm{fit}\left(\frac{k}{k_*}\right)
^{n_\mathrm{fit}+\frac{1}{2}\alpha_\mathrm{fit}\ln \frac{k}{k_*}},
\end{equation}
the fitting parameters being $A_\mathrm{fit}$, $n_\mathrm{fit}$ and
$\alpha_\mathrm{fit}$. The spectral index is evaluated at the pivot
scale: 
\begin{equation}
n_S(k_*)=1+n_\mathrm{fit}.
\end{equation}
The running is parameterized by $\alpha_\mathrm{fit}$.  This fitting
form is not self-consistent since it is not possible to have strictly
constant $n$ and non-zero $\alpha$ given the above definitions.  This
inconsistency manifests itself by the uncontrolled growth of errors
away from the pivot scale.  

In the following we will describe how to obtain approximate solutions
for the scalar and tensor power spectra and the spectral indices and
their running, assuming that the evolution of the background
quantities is known.

\section{The Uniform Approximation}
\label{uniapproximation}

In two recent papers~\cite{hhjm,hhhjm} we have implemented an approach
based on the uniform approximation (for a detailed description of the
approximation, see Ref.~\cite{Olver}) to calculate the power
spectra and the associated spectral indices of primordial
perturbations from inflation. Our method leads to simple expressions
for the power spectra and spectral indices with calculable error
bounds. The solution for the modes $u_k$ can in principle be
determined to arbitrary order in the uniform approximation. In
practice, results accurate to sub-percent level are obtained at
next-to-leading order.

While a detailed derivation of the mode equations, power spectra, and
spectral indices with the corresponding error terms is given in
Ref.~\cite{hhhjm}, for completeness and to set notation, we provide a
brief summary of the main equations.  In addition, we give an
expression for the scalar to tensor power spectrum ratio $R(k)$ with
the corresponding error term and explain how to compute the error
terms in the uniform approximation in an efficient way.  A new result,
shown here, is that a simple improvement of the leading order uniform
approximation leads to very good accuracy, generically better than
$0.1\%$. This improvement is obtained by utilizing knowledge of the
next-to-leading order results without actually implementing them
fully.

\subsection{The Mode Functions}
\label{modes}
In order to solve for the scalar and tensor mode functions in the
uniform approximation, it is necessary to rewrite the differential
equations~(\ref{modeu}) and (\ref{modev}) in the following form:
\begin{equation}
u_k^{\prime\prime}(\eta)=\left\{-k^2+{1\over\eta^2} 
\left[\nu_S^2(\eta)-{1\over 4}\right]\right\}u_k(\eta),
\label{unu}
\end{equation}
where $\nu_S^2=(z''/z)\eta^2 + 1/4$, and
\begin{equation}
v_k^{\prime\prime}(\eta)=\left\{-k^2+{1\over\eta^2} 
\left[\nu_T^2(\eta)-{1\over 4}\right]\right\}v_k(\eta),
\end{equation}
where $\nu_T^2=(a''/a)\eta^2 + 1/4$.  The shift of $1/4$ in the
definition of $\nu_S^2$ and $\nu_T^2$ is necessary in order to have a
convergent error control function~\cite{hhhjm}. In the following we
will describe the analysis for the scalar modes and present only the
final results for the tensor modes.

The general solution for $u_k(\eta)$ is a linear combination of the
two fundamental solutions $u_k^{(1)}(\eta)$ and $u_k^{(2)}(\eta)$,
viz.,
\begin{equation}
u_k(\eta)=A(k)u_k^{(1)}(\eta) + B(k)u_k^{(2)}(\eta), 
\end{equation}
independent of the order of the approximation.  To fix the coefficients
$A(k)$ and $B(k)$, a linear combination of $u^{(1)}_k(\eta)$ and
$u^{(2)}_k(\eta)$ must be taken so that
$u_k(\eta)=e^{-ik\eta}/\sqrt{2k}$ in the limit
$k\eta\rightarrow-\infty$.  With proper normalization, the solution
for $u_k$ is, in leading order,
\begin{equation}
u_{k,1,\lgtr}(\eta)=\sqrt{\frac{\pi}{2}}Cf_{S,\lgtr}^{1/4}(k,\eta)
g_S^{-1/4}(k,\eta)\left[{\rm  Ai}(f_{\lgtr})-i{\rm Bi}(f_{\lgtr})\right],
\label{ukfull}
\end{equation}
with
\begin{eqnarray}
\label{ff}
f_{S,\lgtr}(k,\eta)&=&\mp\left\{\pm\frac 3
2\int_\eta^{\bar\eta_S}d \eta'\left[\mp 
g_S(k,\eta^{\prime})\right]^{1/2}\right\}^{2/3}\hspace{-3mm},\\
\label{gs}
g_S(k,\eta)&=&\frac{\nu_S^2(\eta)}{\eta^2}-k^2.
\end{eqnarray}
One part of the solution is valid to the left of the turning point
\(\bar\eta_S\), defined as the solution to
\(k^2=\nu_S^2({\bar\eta}_S)/{\bar\eta}_S^2\), and the other part is
valid to the right of the turning point ($\eta\ge\bar\eta$).  The
uniform approximation allows us to calculate bounds on the errors.  We
write
\begin{equation}
u_{k,\lgtr}(\eta)=u_{k,1,\lgtr}(\eta)\left[1+\epsilon_{k,1,\lgtr}(\eta)\right],    
\end{equation}    
where the error term encapsulates the contribution to $u_{k,1,\lgtr}$
beyond leading order. As derived in detail in Ref.~\cite{hhhjm}, the error 
term is bounded by
\begin{eqnarray}
|\epsilon_{k,1,\lgtr}(\eta)|&\le&
\frac{\sqrt{2}}{\lambda}
\left\{\right.
\left[\exp (\lambda{\cal V}_{\eta,\alpha}({\cal
E}))-1\right]\nonumber\\
&&+\left.\left[\exp (\lambda{\cal V}_{\beta,\eta}({\cal E}))-1\right]
\right\},\label{eq:errorbound-u}
\end{eqnarray}
where $\mathcal{V}(\mathcal{E})$ denotes the total variation of the
error control function $\mathcal{E}(\eta)$.  A numerical estimate
shows $\lambda\simeq 1.04$~\cite{Olver}.  The error control function
reads
\begin{eqnarray}
\label{ecf}    
{\cal E}(\eta)&=&\int_{\bar\eta_S}^\eta\left\{
\frac{1}{|g_S|^{1/4}}\frac{d^2}{d\eta^{'2}}
\left(\frac{1}{|g_S|^{1/4}}\right)\right.\nonumber\\
&&\hspace{0.7cm}\left.+\frac{1}{4\eta^{'2}|g_S|^{1/2}}
-\frac{5|g_S|^{1/2}}{16|f_{S,\lgtr}|^3}
\right\}d\eta'.
\end{eqnarray}

The $k\eta\rightarrow 0^-$ limit defines the region of interest for
calculating power spectra and the associated spectral indices.  In
this region, the $1/\eta^2$ pole dominates the behavior of the
solutions and the Airy solution goes over to the LG form leading to
simple expressions for the spectral indices~\cite{hhhjm}. The region
of interest lies to the right of the turning point where the argument
of the Airy functions becomes large.  This allows the approximation of
the Airy functions in terms of exponentials, leading to
\begin{eqnarray}
u_{k,1,>}(\eta)&=&{C\over\sqrt{2}}
g_S^{-1/4}(k,\eta) \left[{\frac 1 2}
\exp\left\{-\frac 2 3 \left[f_{S,>}(k,\eta)\right]^{3/2}
\right\}
\right.\nonumber\\
&&\left.-i \exp\left\{
\frac 2 3\left[f_{S,>}(k,\eta)\right]^{3/2}
\right\}
\right].
\end{eqnarray}
For computing the power spectra in the $k\eta\rightarrow 0^-$ limit, only
the growing part of the solution is relevant:
\begin{equation}
u_{k,1,>}(\eta) =\lim_{k\eta\rightarrow 0^-}
-i C\sqrt{{-\eta\over 2\nu_S(\eta)}}
\exp\left\{
\frac 2 3\left[f_{S,>}(k,\eta)\right]^{3/2}
\right\}.\label{lguk}
\end{equation} 

In a similar fashion the results for the tensor modes $v_k$ can be derived. 
We have
\begin{equation}
v_{k,1,>}(\eta) =\lim_{k\eta\rightarrow 0^-}
-i C\sqrt{{-\eta\over 2\nu_T(\eta)}}
\exp\left\{
\frac 2 3\left[f_{T,>}(k,\eta)\right]^{3/2}
\right\},\label{lgvk}
\end{equation} 
where $f_{T,>}$ indicates that we replace $g_S(k,\eta)$ with
$g_T(k,\eta)$ and $\bar\eta_S$ with $\bar\eta_T$ in Eqn.~(\ref{ff}).

Having obtained expressions for the scalar perturbations $u_k$ and the
tensor perturbations $v_k$ we can now derive the corresponding scalar
and tensor power spectra.

\subsection{The Power Spectra}
\label{powspec}

The expression for the scalar power spectrum $P_S(k)$ as defined in
Eqn.~(\ref{PS}) in the uniform approximation with the corresponding
error term is
\begin{eqnarray}
P_S(k)&=&
\lim_{k\eta\rightarrow 0^-}
\frac{k^3}{2\pi^2}\left|\frac{u_{k,1,>}(\eta)}{z(\eta)}\right|^2\left|1+
\epsilon_{k,1,>}(\eta)\right|^2\nonumber\\
&=& 
\lim_{k\eta\rightarrow 0^-}
P_{1,S}(k)\left[
1+\epsilon_{k,1,S}^P(\eta)\right],
\label{Psp}
\end{eqnarray}
with
\begin{equation}
\label{scalerr}
\epsilon_{k,1,S}^P = 2\mathrm{Re}~\epsilon_{k,1,>} + |\epsilon_{k,1,>}|^2,
\end{equation}
where $P_{1,S}(k)$ denotes the power spectrum for the scalar
perturbations in the leading order approximation.  We can now
substitute either the full expression for $u_k$ given in
Eqn.~(\ref{ukfull}) or the LG form from Eqn.~(\ref{lguk}).

Using the LG expression for $u_k$ we have 
\begin{equation}\label{PSP1}
P_{1,S}(k)=\lim_{k\eta\rightarrow 0^-}
\frac{k^3}{4\pi^2}\frac{1}{|z(\eta)|^2}
\frac{-\eta}{\nu_S(\eta)}
\exp\left\{
\frac 4 3 \left[f_{S,>}(k,\eta)\right]^{3/2}
\right\},
\end{equation}
with the error term for the power spectrum given in
Eqn.~(\ref{scalerr}). The calculation for the tensor power spectrum
follows along the same lines, yielding
\begin{equation}\label{PTP1}
P_{1,T}(k)=
\lim_{k\eta\rightarrow 0^-}
\frac{k^3}{4\pi^2}\frac{1}{|a(\eta)|^2}
\frac{-\eta}{\nu_T(\eta)}
\exp\left\{
\frac 4 3\left[{f}_{T,>}(k,\eta)\right]^{3/2}
\right\},
\end{equation}
with the error term in the same form as in Eqn.~(\ref{scalerr}) with
the substitution $S\rightarrow T$.

The tensor to scalar ratio, $R(k)$, is given by
\begin{equation}
R(k)=\frac{8P_{1,T}(k)}{P_{1,S}(k)}\left(1+\epsilon_{k,1}^{R}\right),
\end{equation}
with the error term
\begin{equation}
\epsilon_{k,1}^{R}=\frac{1+\epsilon_{k,1,T}^P}{1
+\epsilon_{k,1,S}^P}-1.
\end{equation}
In the case of power-law inflation, where $\nu$ is constant, the error
is identically zero, as in this case $\nu_S=\nu_T$. In other words,
the ratio of tensor to scalar perturbations for power-law inflation is
exact already at leading order in the uniform approximation.

\subsection{Estimate of the Error Bound}

Although we can calculate the error bound for the
power spectra from the general expressions in
Eqns.~(\ref{eq:errorbound-u}) and (\ref{ecf}), it is convenient to
have simpler estimates for the errors.

We begin by considering the case of constant $\nu$, where the
$k$-independent error bound for the power spectrum in leading order of
the uniform approximation is~\cite{Olver}
\begin{equation}
|\epsilon_{1}^P|\le 2\sqrt{2}\left(\frac{1}{6\nu}+\frac{\lambda}{72\nu^2}+
\frac{1}{36\sqrt{2}\nu^2}+\cdots \right),
\end{equation}
the generic $\nu$ denoting either of $\nu_S$ or $\nu_T$.
This bound is rigorous and useful, though the prefactor
is not optimally sharp for the case of constant $\nu$.

Suppose now that $\nu(\eta)$ varies slowly with time.
Fix \(k\) and consider the value of \(\nu(\eta)\) at the
turning point \(\bar\eta(k)\), defined by \(k\bar\eta=-\nu(\bar\eta)\).
Given the slow variation of \(\nu(\eta)\), this value \(\bar\nu(k)\)
is a slowly varying function of \(k\).
By expanding the expression for the error control function
of Eqn.~(\ref{ecf})
locally around the turning point, we obtain what is in effect
a derivative expansion for the error term. The leading
term in this expansion, which is free of derivatives,
has the same form as the expression above for constant \(\nu\),
though it now carries the mild \(k\)-dependence of the
variable \(\nu\) case,
\begin{equation}
|\epsilon_{k,1}^P|\le 2\sqrt{2}\left[\frac{1}{6\bar{\nu}(k)}
+\frac{\lambda}{72\bar{\nu}^2(k)}+
\frac{1}{36\sqrt{2}\bar{\nu}^2(k)}+\cdots \right].
\label{eq:errorkdep}
\end{equation}
This bound is not meant to be rigorous, since higher order terms
in the derivative expansion are not included. However, it is
effective and useful in the case of slowly varying \(\nu(\eta)\).
For further discussion of local approximations of this type,
see also Section~\ref{localapp}. In Ref.~\cite{hhhjm} we have
extensively discussed how a non-constant $\nu$ gives rise to a
non-vanishing and $k$-dependent error for the spectral index.

\subsection{Improvement of the Leading Order Result}
\label{improve}

In this section we present a new improvement for the leading-order
uniform approximation.  In previous work~\cite{hhhjm} we derived
next-to-leading order results for $u_k$ and $v_k$ and the
corresponding power spectra and spectral indices.  The resulting
expressions contained several integrals, which are tedious to
evaluate.  On the other hand, the results at next-to-leading order for
the case of constant $\nu$ turned out to be very simple. In essence,
higher order terms in the uniform approximation generate a
prefactor which occurs in the Stirling series for
$\Gamma(\nu)$~\cite{temme}
\begin{equation}
\Gamma^*(\nu)\equiv 1+\frac{1}{12\nu}+\frac{1}{288\nu^2}-\frac{139}{51840\nu^3}
+\cdots,
\end{equation} 
which we know a-priori to be present in the case of constant $\nu$. 
This all-orders prefactor improves the normalization of the power
spectrum dramatically.

The natural question arises as to whether or not it is possible to
utilize these results to improve the leading order expressions without
recourse to full computation of the sub-leading approximations. For
most viable inflationary models, $\nu$ varies slowly and corrections
from the derivatives of $\nu$ are sub-dominant. The full
next-to-leading order machinery may not be required if $\nu$ is
sufficiently well-behaved. In this section we implement this idea and
derive improved leading-order results for the power spectra.  Note
that, at leading order, the spectral index is exact for constant
$\nu$, thus the main improvement to be expected is in the amplitude of
the power spectrum.

We begin with an argument from our previous work~\cite{hhhjm}, which
was used to understand why the leading-order result for the spectral
index was much more precise than estimated bounds for the power
spectrum such as given in Eqn.~(\ref{eq:errorkdep}) would seem to
indicate.  The error is in fact dominated by an amplitude prefactor
which has only a subdominant contribution to the spectral index.
We split the scattering potential in the form
\begin{equation}\label{epssplit}
  \nu^2(\eta)-\frac{1}{4}=\bar\nu^2-\frac{1}{4}+\nu^2(\eta)-\bar\nu^2,
\end{equation}
and choose the \(\eta\)-independent but \(k\)-dependent constant
\(\bar\nu(k)\) to be the value of \(\nu(\eta)\) at the turning point.
As argued previously~\cite{hhhjm}, this splitting allows us to
identify two separate contributions to the total error term for the
power spectrum,
\begin{equation}
  \epsilon^P_{k,1} = \bar\epsilon + \tilde\epsilon.
\end{equation}
The term \(\bar\epsilon\) arises solely from the ultra-local
contribution in the derivative expansion of \(\nu\), and by explicit
calculation it is known to be of the form
\begin{eqnarray}\label{epsbar}
\bar\epsilon&=&[\Gamma^*(\bar\nu)]^2-1\\
  &=&\frac{1}{6\bar{\nu}(k)}
                 +\frac{1}{72\bar{\nu}^2(k)} -
                 \frac{31}{6480\bar{\nu}^3(k)}
                 -\frac{139}{155520\bar{\nu}^4(k)} + \cdots .\nonumber  
\end{eqnarray}
The remaining error term satisfies an integral equation of the form
considered by Olver~\cite{Olver}, with a reduced inhomogeneity;
explicit calculation leads to the rigorous bound
\begin{equation}
  |\tilde\epsilon| \le
    \frac{\mathcal{V}(\mathcal{E} 
     - \bar\mathcal{E})}{\mathcal{V}(\bar\mathcal{E})}
    \left[1 + \mathcal{O}(|\bar\epsilon|)\right]
    |\bar\epsilon|,
\end{equation}
where \(\mathcal{E}\) is the full error control function,
\(\bar\mathcal{E}\) is the error control function for the case
of constant \(\nu = \bar\nu(k)\), and \(\mathcal{V}(\cdot)\) indicates
total variation as before.

In the generic case of slowly varying \(\nu(\eta)\) this expression
clearly shows that \(\tilde\epsilon\) is significantly reduced in
comparison to \(\bar\epsilon\).  Therefore we are motivated to absorb
the ultra-local contributions from higher order corrections into the
power spectrum, leading to an improved first-order expression
\begin{equation}
\tilde{P}_{1}(k) = P_{1}(k)[\Gamma^*(\bar\nu)]^2.
\label{eq:PSimproved}
\end{equation}
This improvement applies to both the scalar and tensor power spectra,
with \(\nu_S(\eta)\) and \(\nu_T(\eta)\) respectively. Since typical
values of $\nu\sim 2$ occur for both scalar and tensor fluctuations,
the terms on the right hand side of Eqn.~(\ref{eq:PSimproved})
correspond to amplitude corrections of order $10\%$, $0.35\%$,
$0.06\%$, and $0.006\%$, respectively. These can be viewed primarily
as local normalization corrections.

To complete the discussion, we can obtain a non-rigorous estimate for
the size of \(\tilde\epsilon\), again using a derivative expansion in
\(\mathcal{E}\) to isolate the leading local contributions. In this
expansion, the leading derivative-free terms in \(\mathcal{E} -
\bar\mathcal{E}\) must cancel, and therefore the leading term is
proportional to the first derivative of \(\nu(\eta)\). After some
calculation we find
\begin{eqnarray}
\label{errorthird}
  |\tilde\epsilon|
    &\le&
        \frac{3}{2} \left|\frac{\bar\nu'}{k}\right|
        \left[1 + \mathcal{O}(|\bar\epsilon|)\right]
        |\bar\epsilon| \nonumber \\
    &\le&
        \frac{1}{4\bar\nu(k)} \left| \frac{d \ln \bar\nu(k)}{d \ln k} \right|
        \left[1 + \mathcal{O}\left(\frac{1}{\bar\nu(k)}\right)\right],
\end{eqnarray}
where we have used the chain rule to write the derivative in terms of
a derivative of \(\bar\nu(k)\) with respect to \(k\) and have used the
explicit form for \(\bar\epsilon(k)\).  It is understood that this is
not a rigorous inequality, since we have neglected the higher order
terms in the derivative expansion, but it is well-motivated and useful
for slowly-varying \(\nu(\eta)\). This error estimate determines the
order to which the local corrections~(\ref{eq:PSimproved}) should be
taken into account. For instance, when $\nu$ is slowly varying, most
of the error can be compensated using Eqn.~(\ref{eq:PSimproved}).
Examples will be encountered in Section~\ref{examples}, where we
explicitly demonstrate the success of this improvement procedure. As
is to be expected, the ratio $R(k)$ is much less sensitive to the
normalization error compared to the amplitude of the power spectrum.
For this quantity, the error is well-estimated by Eqn.~(\ref{errorthird}).

\subsection{\label{specind}The Spectral Indices}

We now turn to the evaluation of spectral indices and corresponding
error terms from the power spectra computed using the uniform
approximation.  We will discuss further simplifications leading to
local expressions for $n_S$ and $n_T$.

\subsubsection{Integral Expression for the Spectral Index}
\label{specint}

The differentiation of the power spectrum with respect to $k$ is
straightforward.  As stated earlier, it is important to
remember that the turning point $\bar\eta_S$ is a function of $k$
since $k=-\nu_S(\bar\eta_S)/|\bar\eta_S|$ where $\nu_S(\bar\eta_S)$ is
the value of $\nu_S(\eta)$ at the turning point
$\eta=\bar\eta_S$. Using this relation, we find
\begin{equation}
n_{1,S}(k)=4-2k^2
\lim_{k\eta\rightarrow 0^-}\int_{\bar\eta_S}^\eta\frac{d\eta'}
{\sqrt{g_S(k,\eta')}}.\label{nsint1}
\end{equation}
The error for the spectral index only arises from the $k$-dependent
part of the error in the power spectrum.  Therefore, the error in the
spectral index is sensitive only to the time variation of $\nu_S$.  To
estimate this error, the spectral index as written in
Eqn.~(\ref{defns}) can be expressed via the leading order power
spectrum in the following form:
\begin{eqnarray}
n_S(k)&\simeq&1+\frac{d\ln P_{1,S}}{d\ln k}
+k\frac{d\epsilon_{k,1,S}^P}{d k}\nonumber\\
&\equiv&n_{1,S}(k)+\epsilon_{k,1,S}^n,
\end{eqnarray}
with
\begin{equation}
\label{errorn}
\epsilon_{k,1,S}^n \equiv {k}\frac{d\epsilon_{k,1,S}^P}{d k},
\end{equation}
where we estimate $\epsilon_{k,1,S}^P$ by the leading order error term
in Eqn.~(\ref{epssplit}), i.e., by $\bar \epsilon$ as given in
Eqn.~(\ref{epsbar}). It should be noted that the leading errors in the
spectral index are proportional to $k$ derivatives of $\bar{\nu}(k)$,
unlike the situation for the power spectrum [Eqn.~(\ref{epsbar})]. 

The above analysis can be carried out for tensor perturbations in an
identical fashion, including the error estimation, with the
replacement $\nu_S\rightarrow\nu_T$.  The spectral index for
gravitational waves is given by 
\begin{equation} 
n_{1,T}(k)=3-2k^2
\lim_{k\eta\rightarrow 0^-}
\int_{\bar\eta_T}^\eta\frac{d\eta'} {\sqrt{g_T(k,\eta')}}.
\label{ngint}
\end{equation}

\subsubsection{The Local Approximation}
\label{localapp}

It is possible to simplify the expressions for the spectral indices as
given in Eqns.~(\ref{nsint1}) and (\ref{ngint}) by approximating the
integrals further.  By doing so, we lose the ability to quantify the
error estimates for $n_S$ and $n_T$ but are able to write down local
expressions for the spectral indices. The integrands in
Eqns.~(\ref{nsint1}) and (\ref{ngint}) have square-root singularities
at the turning points, i.e., at the lower integration limits.  At the
upper limit, $\eta$ goes to zero and the integrands vanish linearly,
therefore, assuming $\nu_S(\eta)$ and $\nu_T(\eta)$, respectively, are
well-behaved, we expect the main contribution to the integrals to
arise from the lower limit.  Using the knowledge that $\nu_S$ and
$\nu_T$ are slowly varying, we expand them around their turning points
in Taylor series.  To second order in derivatives,
\begin{equation}
\nu_S^2(\eta)\simeq
\bar\nu_S^2+2\bar\nu_S\bar\nu'_S
\left(\eta-\bar\eta_S\right)
+({\bar\nu_S}^{\prime 2}+\bar\nu_S''\bar\nu_S)(\eta-\bar\eta_S)^2, 
\label{nuexp}
\end{equation}
where a bar indicates that the quantity has to be evaluated at the
turning point. For $\nu_T$ we obtain a similar expression with the 
index $S$ replaced by $T$. We can now solve the integrals in
Eqns.~(\ref{nsint1}) and (\ref{ngint}) exactly and find for the scalar
spectral index 
\begin{eqnarray}
\label{nslocal}
n_S(k)&\simeq&4-2\bar\nu_S\left\{1-\frac{\bar\nu'_S}{\bar\nu_S}\bar\eta_S
\left(1-\frac\pi 2\right)\right.\\
&&\left.+\frac{\bar\eta^2_S}{2}\left[
\frac{\bar\nu'^{2}_S}{\bar\nu^2_S}(2-\pi)
+\frac{\bar\nu''_S}{\bar\nu_S}(1-\pi)\right]\right\},\nonumber
\end{eqnarray}
and for the tensor spectral index analogously
\begin{eqnarray}
\label{ntlocal}
n_T(k)&\simeq&3-2\bar\nu_T\left\{1-\frac{\bar\nu'_T}{\bar\nu_T}\bar\eta_T
\left(1-\frac\pi 2\right)\right.\\
&&\left.+\frac{\bar\eta^2_T}{2}\left[
\frac{\bar\nu'^{2}_T}{\bar\nu^2_T}(2-\pi)
+\frac{\bar\nu''_T}{\bar\nu_T}(1-\pi)\right]\right\}.\nonumber
\end{eqnarray}

The local approximation is a further simplification of the leading
order result in the uniform approximation and we will test its
accuracy with three numerical examples in Section~\ref{examples}.
We denote the leading contribution of unity in the curly brackets in
Eqns.~(\ref{nslocal}) and (\ref{ntlocal}) as zeroth
order, the second term proportional to $\bar\nu'$ as first order in
the local approximation, and the remaining terms proportional to
$\nu'^2$ and $\nu''$ as second order.

\subsubsection{Spectral Indices in Terms of Slow-Roll Parameters}

It is possible to simplify the expressions for the spectral indices
one step further and obtain results similar to the ones obtained in
the slow-roll approximation to be discussed in the next section.  We
have investigated this further approximation in detail
previously~\cite{hhhjm} and pointed out its shortcomings. Here the
results are summarized for leading order only.

We begin by considering the spectral index for the scalar
perturbations as given in Eqn.~(\ref{nslocal}).  In order to write
$n_S$ in terms of slow-roll parameters which are defined as (see
e.g., Ref.~\cite{sg})
\begin{equation}
\epsilon\equiv-{\dot{H}\over H^2}=\frac 1
2\left(\frac{\dot\phi}{H}\right)^2, 
~\delta_n\equiv \frac{1}{H^n\dot\phi}\frac{d^{n+1}\phi}{dt^{n+1}},
\label{epsdelxi}
\end{equation}
we expand $\bar\nu_S$, $\bar\nu'_S \bar\eta_S$, and $\bar\nu''_S
\bar\eta_S^2$ up to second order in these parameters.  The
contribution $\bar\nu_S'^{2}\bar\eta_S^2/\bar\nu_S$ is already of
fourth order and is neglected.  Using the expression for conformal
time in slow-roll parameters given by
\begin{eqnarray}
\eta&\simeq&-{1\over 
aH}\left(1+\epsilon+3\epsilon^2+2\epsilon\delta_1+\cdots\right),
\label{etasr}
\end{eqnarray}
the relation, $\nu_S^2=(z''/z)\eta^2+1/4$, and 
\begin{equation}
{z^{\prime\prime}\over z}=2a^2H^2\left(1+\epsilon+{3\over 2}\delta_1
+2\epsilon\delta_1+\epsilon^2+\frac 1 2 \delta_2\right),
\label{zdpz} 
\end{equation}
it is easy to write down $\bar\nu$ and its derivatives in terms
$\bar\epsilon$, $\bar\delta_1$, and $\bar\delta_2$ (For a detailed
derivation and explicit expressions, see Ref.~\cite{hhhjm}).  As in
earlier sections, the bar indicates that the  parameters are
to be calculated at the turning point.  Inserting all the expressions
into the local approximation for the scalar spectral index,
Eqn.~(\ref{nslocal}), allows us to write the spectral index in terms of
slow-roll parameters:
\begin{eqnarray}\label{nssrtp}
n_S(k)&\simeq& 1-4\bar{\epsilon}-2\bar{\delta}_1-
8\bar{\epsilon}^2\left(\frac{17}{6}-\pi\right)\\
&&-10\bar{\epsilon}\bar{\delta}_1\left(\frac{73}{30}-\pi\right)
+2(\bar{\delta}_1^2-\bar{\delta}_2)\left(\frac{11}{6}-\pi\right).
\nonumber
\end{eqnarray}
Equivalently, for the tensor spectral index derived from a slow-roll
expansion of the local result, Eqn.~(\ref{ntlocal}), we find
\begin{equation}\label{ntsrtp}
n_T(k)\simeq -2\bar\epsilon-\left(\frac{34}{3}-3\pi\right)\bar\epsilon^2
-\left(\frac{28}{3}-3\pi\right)
\bar\epsilon\bar\delta_1,
\end{equation}
where we have used
\begin{equation}
{a^{\prime\prime}\over a}=2a^2H^2\left(1-{1\over
2}\epsilon\right). \label{adpa} 
\end{equation}
Compared to results given previously~\cite{hhjm, hhhjm}, the
expressions for the spectral indices given here have been fully
expanded to second order in slow-roll parameters.

\section{The Slow-Roll Approach}
\label{slowroll}
The most popular analytic method to evaluate primordial power spectra
and spectral indices is the so-called slow-roll approach. The basic
idea behind this approach is the following. Recall the expression
for $z''/z$ in terms of $\epsilon$, $\delta_1$ and $\delta_2$ as given
in Eqn.~(\ref{zdpz}). Assume now that $\epsilon$ and $\delta_1$ are
small and constant. If this is the case, all terms $\delta_{n>1}$
vanish, since they can be written as derivatives of $\epsilon$ and
$\delta_1$, and all terms of order ${\cal O}(\epsilon^2)$ can be
neglected. It is then also possible to express $a^2H^2$ as
$(1+2\epsilon)/\eta^2$, leading to a simplified equation for $u_k$:
\begin{equation}
u_k''+\left(k^2-A/\eta^2\right)u_k=0, 
\end{equation}
with
\begin{equation}
A=2+6\epsilon+3\delta_1={\rm constant}.
\end{equation}
This equation is immediately solved in terms of Bessel functions. 
Since the constant $A$ will be different for different values of $k$
(as $\epsilon$ and $\delta_1$ are approximated by different constants
for different $k$), for every momentum $k$ a ``best-fit power-law''
result is obtained and expressions for the power spectrum $P_S(k)$ and
the spectral index $n_S(k)$ can be derived.  Tensor
perturbations can be treated in the same way, leading to expressions
for $P_T(k)$ and $n_T(k)$.  The terminology ``slow-roll'' is easy to
understand: the requirement on $\epsilon$ and $\delta_1$ being small
and roughly constant admits inflationary models with smooth
potentials, leading to a phase where the inflaton field rolls slowly
down the potential before inflation ends and reheating begins.  The
shortcoming of the slow-roll approach is that there is no simple way
to improve the approximation beyond leading order.  Once the
assumption that the slow-roll parameters are constant is given up,
at the next formal order of the approximation, the Bessel solution
is no longer valid~\cite{wms}.  Different ways for improving the
slow-roll approximation have been suggested, e.g.,
\cite{sg,srimprov,CGS,gong}, but they often lead to rather involved
expressions for power spectra and spectral indices and, more
importantly, are not error-controlled.

As the slow-roll approach is convenient, widely used, and for most
common inflationary models expected to be a reasonable approximation,
we will compare slow-roll results against the exact mode-by-mode
integration and the uniform approximation in
Section~\ref{examples}. For completeness, we summarize here the
results for the power spectra and spectral indices following closely
the work of Stewart and Lyth~\cite{sl}.

The power spectra in the slow-roll approximation read
\begin{eqnarray}
P_S(k)&\simeq&\left[1+2(2-\ln 2-b)(2\epsilon+\delta_1)-2\epsilon\right]
\nonumber \\
&&\times\frac{2G}{\pi}\frac{H^4}{\dot\phi^2}\bigg|_{aH=k} \ ,
\label{PSsr}\\ 
P_T(k)&\simeq& \left[1-2(\ln 2+b-1)\epsilon\right]
\frac{2GH^2}{\pi}\bigg|_{aH=k},
\label{PTsr}
\end{eqnarray}
where $b$ is the Euler-Mascheroni constant, $2-\ln 2-b\simeq 0.7296$
and $\ln 2+b -1 \simeq 0.2704$. 

The scalar and tensor spectral indices are given by
\begin{eqnarray}
\label{nssr}
n_S(k)&\simeq& 1-4\epsilon-2\delta_1-2(1+c)\epsilon^2
+\frac{1}{2}(3-5c)\epsilon
\delta_1\nonumber 
\\
&&-\frac{1}{2}(3-c)\delta_1^2+\frac{1}{2}(3-c)\delta_2,\\
\label{ntsr}
n_T(k)&\simeq&-2\epsilon-(3+c)\epsilon^2-(1+c)\epsilon\delta_1,
\end{eqnarray}
and $c\simeq 0.08145$. The expressions $n_S(k)\simeq
1-4\epsilon-2\delta_1$ and $n_T(k)\simeq-2\epsilon$ are
the well-known first-order slow-roll results.

In the numerical calculations for slow-roll presented below, we will
approximate the solutions of the mode equations in terms of Bessel
functions, while solving the background equations without any further
approximations.  Hence, we do not assume the strict slow-roll
conditions that the slow roll parameters $\epsilon$, $\delta_1$, and
$\delta_2$ be constant.  This is not the way the slow-roll approach is
often implemented when estimating inflationary parameters: A further
asymptotic approximation is utilized to solve the background
equations~\cite{salbond,LPB} employing a derivative expansion in the
inflationary potential $V(\phi)$.  Such expansions may provide useful
estimates but do not give the precision required by present and
upcoming CMBR observations.  The reason for the inaccuracy arising
from expansions in the potential, as will be seen in an example, is
that the Taylor expansion in derivatives can often fail.  This is not
the case when the parameters are expressed in terms of the Hubble
parameter and its derivatives.  Our aim here is to determine the
accuracy of the mode solutions themselves, therefore we do not focus
on errors arising from approximations to the background equations.

\section{Numerical Implementations}
\label{numerics}

In this section we describe the numerical implementation of the exact
mode-by-mode integration and the uniform approximation and its
simplifications. Technical details can be found in Appendix~\ref{app}.
Since the uniform approximation is useful both in setting initial
conditions for the exact numerical solution, and as a complete
semi-analytic method in its own right, we discuss it first below.

\subsection{Leading Order Uniform Approximation: Numerical Issues}
\subsubsection{Preliminaries}

We begin by addressing some technical points and introducing conventions.
Since the conformal time is defined only up to a constant, we set it
to zero at the end of inflation. If the given model does not have a
natural end to inflation this point is somewhat arbitrary and we set
$\eta$ to zero typically some number of e-folds after the highest mode
of interest freezes out.

The power spectrum and the spectral indices are to be calculated in
the limit $k\eta\to 0^-$, so that the LG approximation can be used. To
avoid accumulation of numerical error, however, these quantities
should not be calculated directly in this limit. Since the freeze-out
happens soon after the turning point is crossed, the computation is
carried out some e-folds after the turning point for each respective
mode, but well before the end of inflation.  The linear combination of
both solutions to the mode equations, in terms of Ai- and
Bi-functions, is then completely dominated by the exponentially
growing part. We found that carrying out the computations 4 or 5
e-folds after the respective turning point provided sufficient
accuracy.

In all the computations described below, the Ai- and Bi-functions were
calculated with the algorithm given in Ref.~\cite{airynums}.

\subsubsection{The Mode Solutions}

The full solutions $u_{k,1,>}(\eta)$ and $u_{k,1,<}(\eta)$ are not
needed for the calculation of the power spectrum and the spectral
index.  Nevertheless, it is useful to calculate some of them for
selected momenta $k$ in order to compare the leading order uniform
solutions to the exact numerical solutions found from a mode-by-mode
integration of the differential equations.  For $\eta<\bar{\eta}$ we
can calculate the integrals appearing in $f_<(k,\eta)$, as defined in
Eqn.~(\ref{ff}) numerically via
\begin{equation}
\int_\eta^{\bar{\eta}}{d\eta'}\sqrt{g(k,\eta')}
=\left(\int_\eta^{\eta_i}+\int_{\eta_i}^{\bar{\eta}}\right)
{d\eta'}\sqrt{g(k,\eta')},
\end{equation}
where $\eta_i$ is an initial value of the conformal time.  In the
actual numerical routine we therefore have to know the second integral
before we can calculate the uniform solutions left of the turning
point: This is achieved by an additional run of the integrator for
the background equations.
 
\subsubsection{Power Spectra and Spectral Indices}

The power spectra for scalar and tensor perturbations in leading order
of the uniform approximation are given by
Eqns.~(\ref{PSP1})~and~(\ref{PTP1}). The integrals are calculated
using a trapezoidal rule with non-equidistant discretization in
conformal time. As mentioned above, we avoid calculating the spectra
numerically in the limit $k\eta\to 0^-$, but instead do so some 4-5
e-folds after the turning point. In the case of power-law inflation
analytical results are available for the leading order contributions
to the power spectra (see Ref.~\cite{hhhjm}). We have checked that the
power spectra numerically calculated from Eqns.~(\ref{PSP1}) and
(\ref{PTP1}) are in agreement with these analytic results (see
Section~\ref{plinf} and Table~\ref{tab1} for details).
 
The spectral indices in the uniform approximation may be calculated
either by numerically differentiating the power spectrum as described
in Section~\ref{specind} for the exact mode-by-mode approach, or by
using the formulae in Eqns.~(\ref{nsint1}) and (\ref{ngint}).  In
Appendix~\ref{specindapp} we describe how to deal with the inverse
square root singularities appearing in the integrals to be performed
in the second case.

\subsection{Mode-by-Mode Numerical Integration}

\subsubsection{Initial Conditions and Mode Functions}

In order to numerically obtain the mode functions, we must satisfy
the initial condition requirement, i.e., in the limit $k\eta\to
-\infty$,  
\begin{equation}
u_k(\eta)\longrightarrow \frac{1}{\sqrt{2k}}e^{-ik\eta}.
\label{bdvac}
\end{equation}
Two difficulties in imposing this formal initial condition immediately
arise. First, in any numerical solution, the calculation must begin at
a finite initial time, thus for modes with small enough values of $k$,
the condition $k\eta\to -\infty$ may not be fulfilled. Second,
for modes at larger $k$ values, there are very many oscillations
before the turning point is reached, naively requiring very fine
time-steps if the entire temporal range must be handled numerically.

\begin{figure}[t]
  \centering
  \includegraphics[width=0.97\columnwidth]{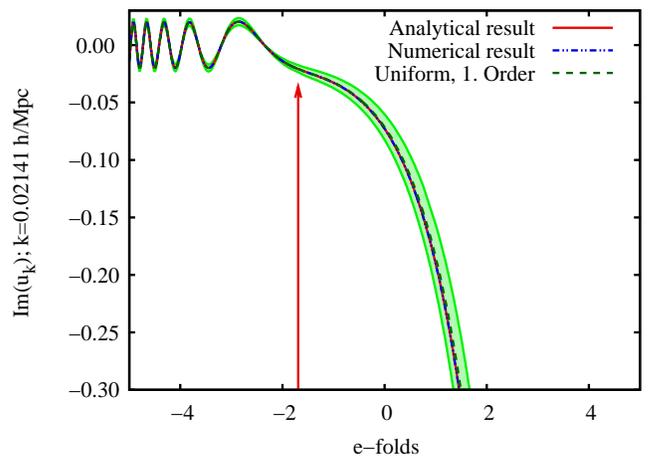}
  \caption{Imaginary (growing) part of the scalar mode function
    for a power-law model with $p=11$ for the mode
    $k=0.0214\,h\,\mathrm{Mpc}^{-1}$. The solid red line is the exact
    analytical solution, the dashed-dotted blue line the result from
    the numerical solution of the exact equations, and the dashed
    green line the uniform approximation to leading order; the
    analytical and uniform approximation results are almost on top of
    each other. The arrow shows the turning point and the green band
    is the estimated error bound for the leading order uniform
    approximation.}
  \label{fig:Imuk-m20.01}
\end{figure}

\begin{figure}[hbtp]
  \centering
  \includegraphics[width=0.97\columnwidth]{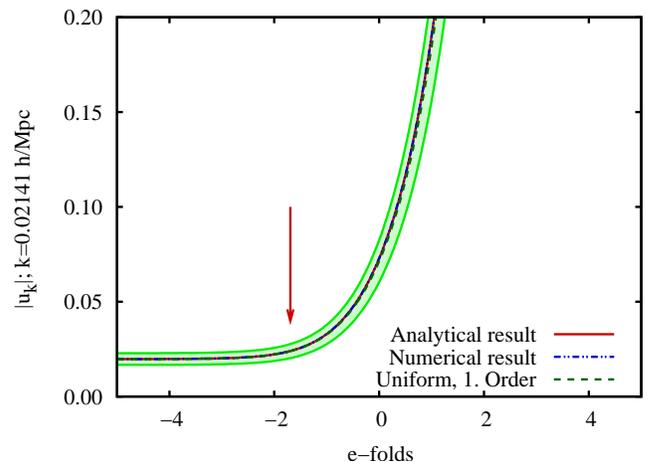}
  \caption{The absolute value of the scalar mode function
    for a power-law model following Fig.~\ref{fig:Imuk-m20.01}.} 
  \label{fig:absuk-m20.01}
\end{figure}

To circumvent these problems, we use the uniform approximation to set
initial conditions in a regime where it is exponentially accurate.
For each mode, we take as initial condition the uniform approximation
result at roughly 20 zeros, i.e., 10 oscillations, before the turning
point for that mode.  The number of zeros from a given time to the
turning point can be estimated by $n\pi\approx k
[\bar{\eta}(k)-\eta]$.  As shown below, this initialization procedure
suppresses numerical errors, especially in the high $k$ regime where
the precision of the power spectrum and the spectral index is improved
without taking smaller and smaller time steps.  In addition, only a
smaller number of ``active'' modes, i.e., the modes that are within
some 20 zeros before the turning point and not yet frozen, need be
considered at any time.  In the regime of small $k$, we first note
that as inflation has to start somewhere in practical numerical
calculations, this introduces a lowest value for $k$, defined by the
criterion that the mode should be well inside the Hubble length.  As
the initial conditions for inflation are unknown, we will assume (i)
that inflation began well before the 55-65 e-folds necessary to solve
the flatness and horizon problems, and (ii) by the time our
calculations are to be performed, Eqn.~(\ref{bdvac}) applies. By
isolating how initial conditions are defined from possible early-time
artifacts, our method of implementing initial conditions also leads to
substantial improvement in the low-$k$ regime as well.

\begin{figure}[thbp]
  \centering
  \includegraphics[width=0.97\columnwidth]{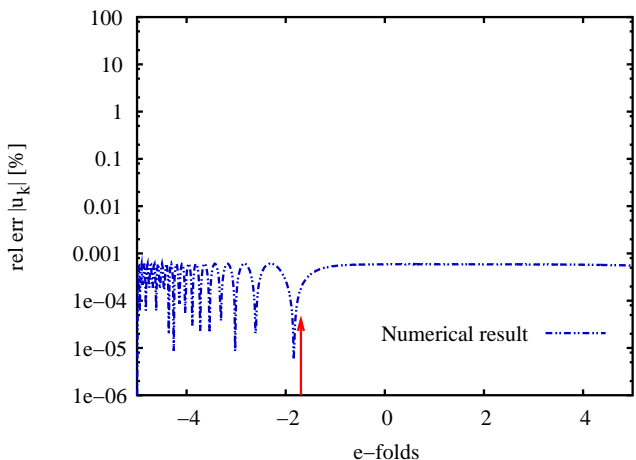}
  \caption{Relative error of $|u_k|$ shown in
    Fig.~\ref{fig:absuk-m20.01} for the numerical calculation (blue
    line). The arrow shows the turning point for the mode.} 
  \label{fig:absdiff-uk}
\end{figure}

\begin{figure}[thbp]
  \centering \includegraphics[width=0.97\columnwidth]{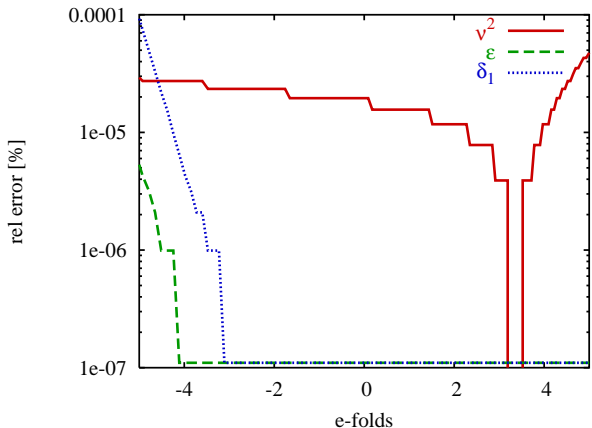}
  \caption{Relative errors of $\nu$, $\epsilon$, and $\delta_1$; 
  model parameters are as specified in Fig.~\ref{fig:Imuk-m20.01}.} 
  \label{fig:nuconst}
\end{figure}

In Figs.~\ref{fig:Imuk-m20.01} and \ref{fig:absuk-m20.01} we display
the imaginary (growing) part and absolute value of $u_k$ for a mode
with $k=0.0214\,h\,\mathrm{Mpc}^{-1}$ for a power-law inflation model. 
As can be seen in both cases, the uniform approximation at leading
order and the numerical results are clearly very close.  The error in
the absolute value $|u_k|$ (relevant to the power spectrum) from the
mode-by-mode integration is shown in Fig.~\ref{fig:absdiff-uk}.  The
numerical error is less than 1 part in $10^5$.  To check the accuracy
of the solution of the background equations, the numerical deviations
from the expected pure constant values for $\nu$, $\epsilon$, and
$\delta_1$ are shown in Fig.~\ref{fig:nuconst}; the errors are
comfortably below parts per million.  Detailed quantitative results
are given in the following section.

\subsubsection{Power Spectra and Spectral Indices}

For the case of scalar perturbations the modes freeze out once the
power spectrum $P_S(k,\eta)$, as defined in Eqn.~(\ref{PS}), becomes
constant as a function of conformal time, i.e., when
$P_S'(k,\eta_\mathrm{freeze})\simeq 0$ at a numerically determined
freeze-out time $\eta_\mathrm{freeze}$.  We found that it is
numerically robust to track the time derivative of $P_S(k,\eta)$ to
determine this freeze-out. The same situation holds for the tensor
perturbations and $P_T(k,\eta)$ as defined in Eqn.~(\ref{PT}).

Once the power spectra have been obtained, the spectral indices for
scalar and tensor perturbations are found by evaluating the
derivatives of the power spectra with respect to $k$, as defined in
Eqns.~(\ref{defns}) and (\ref{defnt}).  The derivatives are computed
numerically with nonequidistant momentum discretization due to the
momentum readjustment described in Appendix~\ref{mom}.  We take three
discretization points and approximate the derivative by a
non-symmetric, second-order, rule.  In the numerical evaluation of the
derivatives, for every individual $k$ we use two very close
neighboring points with $\Delta \ln k \approx 0.01$--$0.02$ over the
entire $k$-range considered.

\section{Results}
\label{examples}

In this section we discuss three inflation models.  We start with
power-law inflation, one of the very few analytically solvable cases. 
Next we investigate two chaotic inflation models, with a quadratic and
quartic potential, respectively.  Finally, we present a model in which
the third derivative in the potential $V(\phi)$ does not exist at one
point, leading to a ``kink'' in $z''/z$.

We present below results from numerical solution of the mode
equations, the uniform approximation with different simplifications
and improvements, and slow-roll approximations in first and second
order.  In all cases, we show the results for the ratio of tensor and
scalar power spectra, as well as the scalar and tensor spectral
indices.  Some of the results are presented in tables, for a specific
value of the momentum $k$, while selected results are shown in
figures.  In addition to the basic quantities themselves, we also give
relative errors in some cases.

\subsection{Power-Law Inflation}
\label{plinf}

\begin{table*}[t]
\caption{\label{tab1}Numerical precision tests in the power-law case
with $p=11$ at $k_*=0.06875 h\mathrm{\, Mpc^{-1}}=0.0495\mathrm{\
Mpc^{-1}}$ ($h=0.72\pm 0.05$; the WMAP pivot scale is at
$0.05\mathrm{\ Mpc^{-1}}$, see Ref.~\cite{spergel}) in the various
approximations.}
\begin{ruledtabular}
\begin{tabular}{lccccc}
Approximation         
& $R(k_*)$ & $n_S(k_*)$ & $\alpha_S(k_*)$ &$n_T(k_*)$ &$\alpha_T(k_*)$\\
\hline
Analytical            
&1.454545& 0.8   & 0          &-0.2   &0          \\
Numerical\footnotemark[1]             
&1.454544&0.79998&$<10^{-7}$&-0.20002&$<10^{-7}$\\
Uniform, 1. order\footnotemark[2]      
&1.454541&0.79992&0.00001&-0.20008&0.00001\\
Local approx., 0. order\footnotemark[2] 
&--      &0.80000&$<10^{-7}$&-0.20000&$<10^{-7}$\\
Local approx., 1. order\footnotemark[2] 
&--      &0.79999&$<10^{-6}$&-0.20001&$<10^{-6}$  \\
Uniform, Slow-Roll redux\footnotemark[3]
&--      &0.80165&$<10^{-7}$&-0.19835&$<10^{-7}$  \\
Slow-Roll, 1. order\footnotemark[4]      
&1.454545&0.81818&$<10^{-7}$&-0.18182&$<10^{-7}$  \\
Slow-Roll, 2. order\footnotemark[5]     
&--      &0.80165&$<10^{-7}$&-0.19835&$<10^{-7}$  \\
\footnotetext[1]{mode-by-mode integration}
\footnotetext[2]{already exact in that order}
\footnotetext[3]{see Eqn.~(\ref{nssrtp}) and Eqn.~(\ref{ntsrtp})}
\footnotetext[4]{expected results: $n_S=1-2/p$ and $n_T=-2/p$}
\footnotetext[5]{expected results: $n_S=1-2/p-2/p^2$ and $n_T=-2/p-2/p^2$ }
\end{tabular}
\end{ruledtabular}
\end{table*}

As one of the few analytically tractable models, power-law
inflation~\cite{MatRatra} provides a useful foil for testing
approximations.  This feature has maintained its popularity, even
though the basic model is not realistic, as inflation never comes to an
end. We compared detailed results from the uniform approximation in
leading and next-to-leading order with slow-roll and exact results
previously~\cite{hhhjm}: The spectral index was exact already in
leading order in the uniform approximation, while the slow-roll
approximation yields a Taylor expansion of the spectral index.  The
amplitude of the power spectrum in the uniform approximation was
accurate to roughly 10\% in leading order while in next-to-leading
order the accuracy of the amplitude improved dramatically with an
error smaller than 1\%.  These results hold for widely differing
values of $p$, the power in the time evolution of the expansion
$a\propto t^p$ [the potential is $V(\phi)=V_0\exp({\sqrt{2/p}\ 
  \phi}$)].  In contrast, the slow-roll approximation being an
expansion in $1/p$, has large errors for small $p$, decreasing as $p$
increases.

Here, we use the power-law model for testing the accuracy of our
numerical implementations of the exact mode equations, as well as the
uniform and slow-roll approximations.  The spectral indices are
constant, therefore the running of the spectral index -- which
measures the $k$-dependence of $n_S$ and $n_T$ -- vanishes.  In
addition, the ratio $R(k)=16/p$ of the power spectra is constant. 
Instead of showing our results graphically -- all curves would be
almost indistinguishable by eye -- we summarize our findings in
Table~\ref{tab1}.  We quote the results for $R(k)$, the spectral
indices, and their running $\alpha_S$ and $\alpha_T$ at the WMAP pivot
scale $k_*=0.0495\mathrm{\ Mpc^{-1}}$~\cite{spergel}.  We have picked
the power $p=11$, which is of course much too small to be considered a
realistic inflationary model, but is perfectly acceptable for
illustrative purposes.  In the uniform approximation there is an
explicit expression for the spectral indices $n_S$ and $n_T$, while
the tensor to scalar ratio $R(k)$ is calculated via the amplitudes of
the power spectra.  In the purely numerical computation, the spectral
index is obtained by taking numerical derivatives of the power
spectrum as as described earlier.

The error in the exact numerical determination of the ratio of the
power spectra $R_*(k)$ is smaller than 1 part in $10^6$, while the
numerical error in the implementation of the uniform approximation is
slightly larger (3 parts in $10^{6}$).  In the case of power-law
inflation, the spectral indices do not accurately reproduce the
expected analytic results, though they do reproduce the expected
slow-roll results to $1/p$-order with the slow-roll parameters being
constant in time (in this case, $\epsilon=-\delta_1=1/p$ and
$\delta_2=2/p^2$). This provides a check on the integration of the
background equations, but has no further useful information.  

For the scalar spectral index $n_S(k_*)$ we find a relative error in
the exact numerical calculation of $0.0025\%$ while for the uniform
approximation the relative error is roughly $0.01\%$.  Again, no error
in the numerical implementation of the slow-roll approximation is
detected.  The situation for the tensor spectral index $n_T(k_*)$ is
similar.  The numerical errors in the uniform approximation for the
spectral indices are slightly larger than for the exact numerical
mode-by-mode approach due to the numerical integration of a function
with an inverse square root singularity.  The results for
$\alpha_S(k_*)$ and $\alpha_T(k_*)$ are very close to zero in all
cases.

In addition, we tested our improvement strategy for the uniform
approximation which should work perfectly in the case of power-law
inflation since $\nu$ is constant.  We found a relative error for
$P_S(k)$ of $0.15\%$ if we include corrections up to $1/\nu^2$, of
$0.025\%$ including corrections up to $1/\nu^3$, and of $0.01\%$ if we
include corrections up to $1/\nu^4$ following
Eqn.~(\ref{eq:PSimproved}).  The corrections are slightly worse than
anticipated but consistent with the small $0.01\%$ error in the
numerical implementation of the uniform approximation.

\subsection{Chaotic Inflation}

\begin{table*}[t]
\caption{\label{tab2}Determination of the characteristic quantities for the 
$\frac{1}{2}m^2\phi^2$ 
chaotic inflation model
at $k_*=0.06875~h\mathrm{\, Mpc^{-1}}=0.0495 \pm 0.0034\mathrm{\ Mpc^{-1}}$ 
($h=0.72\pm 0.05$; the  WMAP pivot scale is at $k=0.05\mathrm{\ Mpc^{-1}}$, 
see Ref~\cite{spergel}) 
in the various approximations; parameters: 
$m^2=(1.89\pm 0.21)\times 10^{-12}/8\pi G$,  
$\phi(0)=16.8/\sqrt{8\pi G}$, 
$\dot\phi(0)=-0.1/\sqrt{8\pi G}/\mathrm{s}$ leading to  
57.655 e-folds of inflation after horizon crossing of $k$=0.0495~Mpc$^{-1}$.}
\begin{ruledtabular}
\begin{tabular}{lccccc}
Approximation         
& $R(k_*)$ & $n_S(k_*)$ & $\alpha_S(k_*)$ &$n_T(k_*)$ &$\alpha_T(k_*)$\\
\hline
Numerical\footnotemark[1]             
&0.13749&0.96507&-0.00064         &-0.01765   &-0.01779          \\
Uniform, 1. order      
&0.13740(14) &0.96505(3)&-0.00064 & -0.01768(1)  &-0.01773 \\
Uniform, improved 1. order\footnotemark[2]      
&0.13749(0)  & 0.96507(0)  &  -0.00064   &-0.01765(0)  &  -0.01773     \\
Local approx., 0. order\footnotemark[3] 
&-- &0.96465 & -0.00066          &-0.01786   &-0.01799          \\
Local approx., 1. order\footnotemark[3] 
&-- &0.96501 & -0.00064          &-0.01761   &-0.01770       \\
Uniform, Slow-Roll redux\footnotemark[4]
&-- &0.96566& -0.00062          &-0.01757   &-0.01768          \\
Slow-Roll, 1. order\footnotemark[5]      
&0.13752 &0.96523 &-0.00063           &-0.01741   &-0.01755       \\
Slow-Roll, 2. order\footnotemark[5]     
&-- &0.96507 &-0.00064           &-0.01764 &-0.01779       \\
\end{tabular}
\end{ruledtabular}
\footnotetext[1]{mode-by-mode integration}
\footnotetext[2]{the improvement here is to second order in powers of
  $1/\nu$, Cf.~Eqn.~(\ref{eq:PSimproved})} 
\footnotetext[3]{see Eqn.~(\ref{nslocal}) and (\ref{ntlocal})}
\footnotetext[4]{see Eqn.~(\ref{nssrtp}) and Eqn.~(\ref{ntsrtp})}
\footnotetext[5]{see Eqn.~(\ref{nssr}) and (\ref{ntsr})}
\end{table*}

\begin{figure*}[htb]
  \centering \vspace{0.5cm}
  \hspace{1cm}(a)\hspace{-1cm}\includegraphics[width=0.97\columnwidth]
  {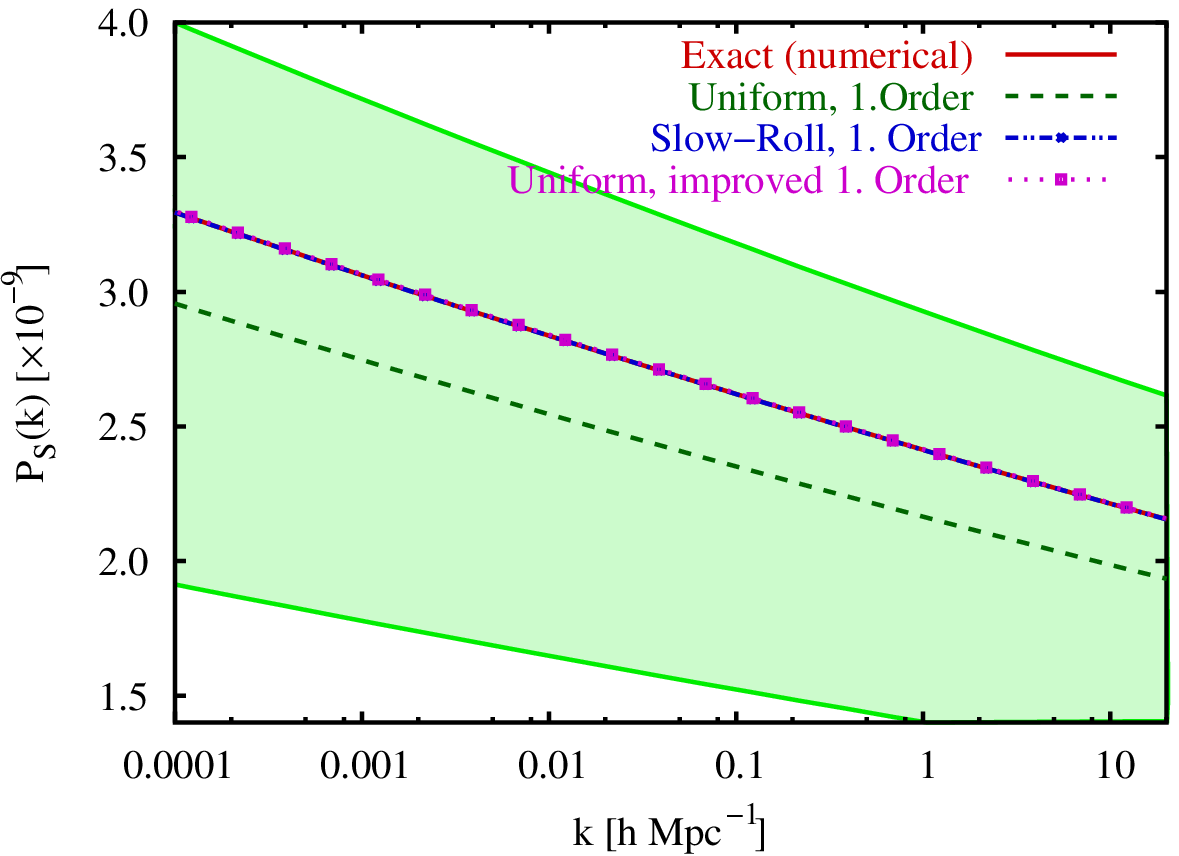}
  \hspace{1cm}(b)\hspace{-1cm}\includegraphics[width=0.97\columnwidth]
  {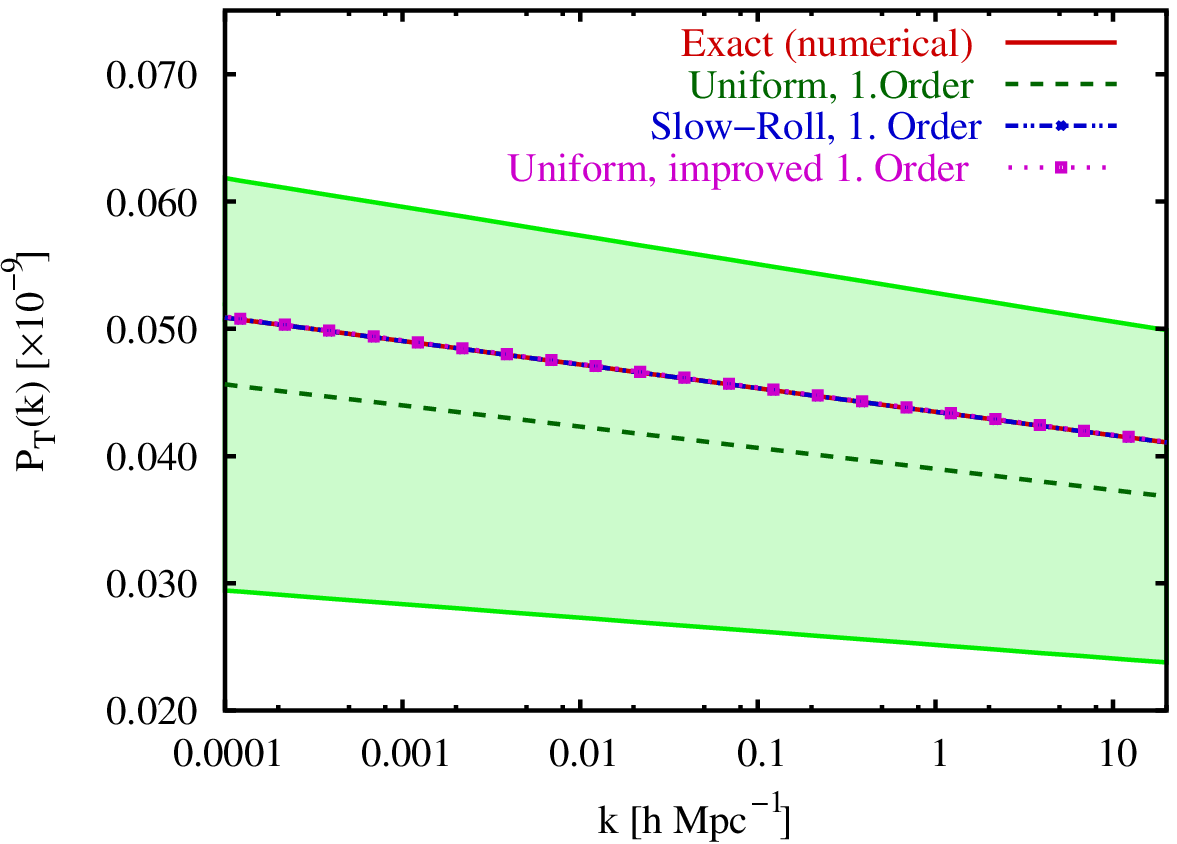} \caption{(a) Scalar power spectrum $P_S(k)$
    and (b) tensor power spectrum $P_T(k)$ for a chaotic
    $\frac{1}{2}m^2\phi^2$-model, parameters as specified in
    Table~\ref{tab2}.  Solid red line: exact numerical results, dashed
    green line: uniform approximation, dashed-dotted blue line:
    slow-roll; dotted purple line: second-order improved uniform
    approximation [Cf. Eqn.~(\ref{eq:PSimproved})]; the green band is
    the estimate for the error bound for the uniform approximation,
    Eqn.~(\ref{eq:errorkdep}).  The exact results and the results from
    the improved uniform and slow-roll approximation are visually on top
    of each other. The error estimate for the improved uniform
    approximation, Eqn.~(\ref{errorthird}), is so small as to be
    indistinguishable from the result itself.  }
  \label{fig:PkS-m20.01}
\end{figure*}

Next we study two chaotic inflation models.  In these examples, the
slow-roll approach is expected to work well; in order to obtain enough
e-folds of inflation the potentials cannot be very steep, thus
slow-roll turns out to be a very good approximation.  For the uniform
approximation we expect similar results as for the power-law case: an
accuracy more or less independent of the chosen parameters.

\subsubsection{Quadratic Potential: $V(\phi)=m^2\phi^2/2$}

\begin{figure*}[t]
  \centering
    \vspace{0.5cm}
    \hspace{1cm}(a)\hspace{-1cm}\includegraphics[width=0.97\columnwidth]
    {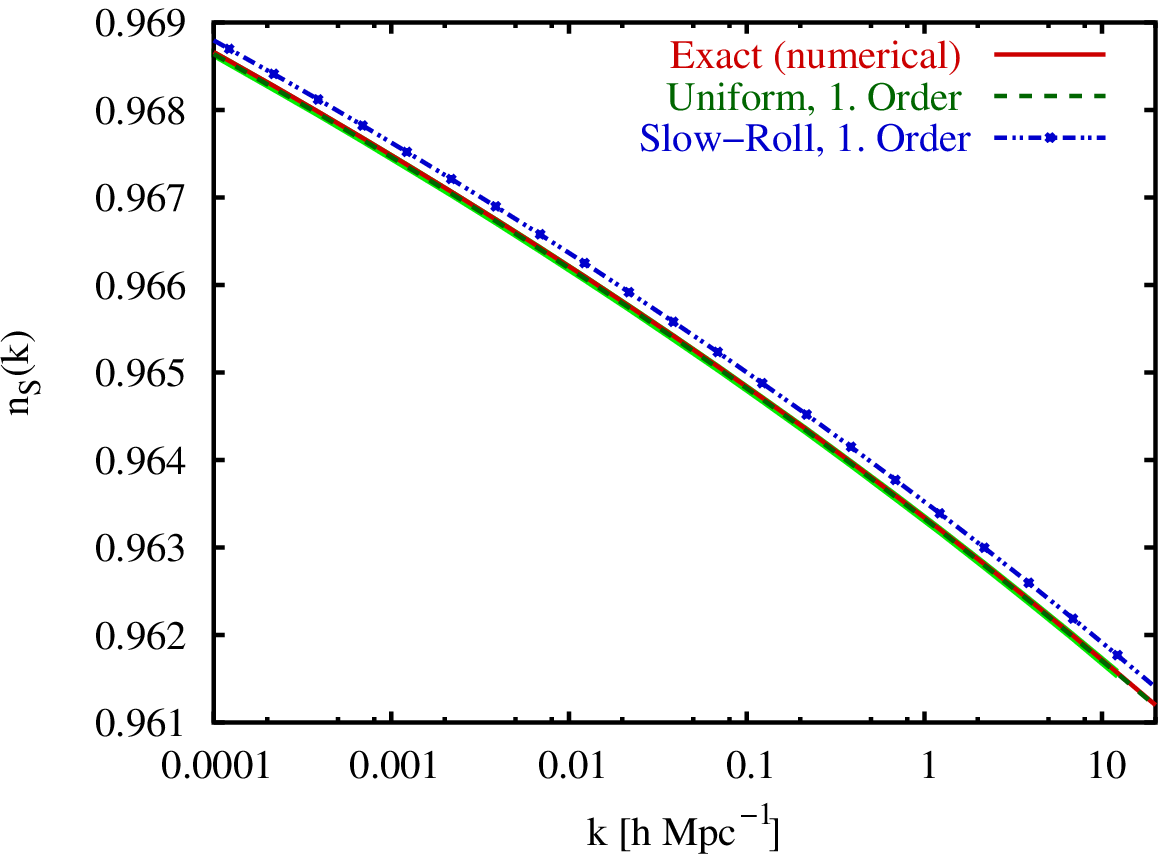}
    \hspace{1cm}(b)\hspace{-1cm}\includegraphics[width=0.97\columnwidth]
    {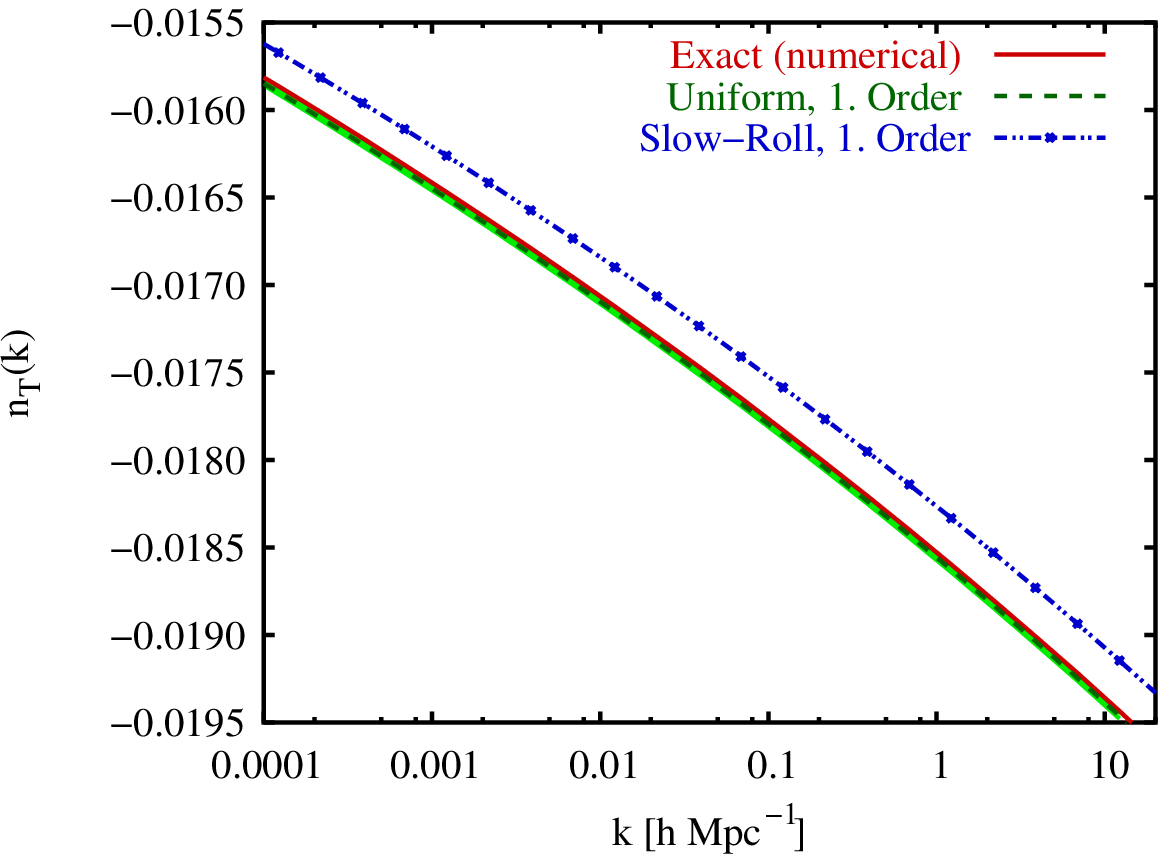} \caption{(a) Scalar spectral index $n_S(k)$
    and (b) tensor spectral index $n_T(k)$ for a chaotic
    $\frac{1}{2}m^2\phi^2$-model, parameters as specified in
    Table~\ref{tab2}.  Solid red line: exact numerical results, dashed
    green line: uniform approximation, dashed-dotted blue line:
    slow-roll; the green band is the error estimate for the uniform
    approximation.  Unlike the case for the power spectrum, accurate
    results for the uniform approximation are obtained without
    recourse to the improvement procedure specified by
    Eqn.~(\ref{eq:PSimproved}).}
  \label{fig:nkS-nkT-m20.01}
\end{figure*}

\begin{figure}[htbp]
  \centering
  \vspace{0.5cm}
  \includegraphics[width=0.97\columnwidth]{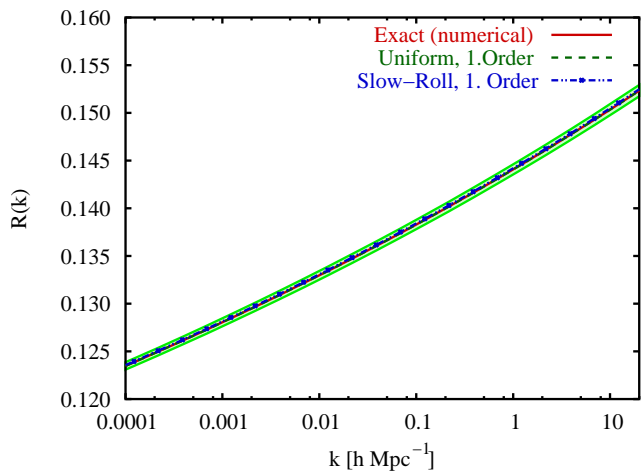}
    \caption{Tensor to scalar ratio $R(k)$ for a chaotic
    $\frac{1}{2}m^2\phi^2$-model; parameters as specified in
    Table~\ref{tab2}.  All three lines lie practically on top of each
    other; see Fig.~\ref{fig:relerror-Rk-m20.01} for the relative
    errors.}
  \label{fig:Rk-m20.01}
\end{figure}

\begin{figure}[htbp]
  \centering
  \vspace{0.5cm}
  \includegraphics[width=0.97\columnwidth]{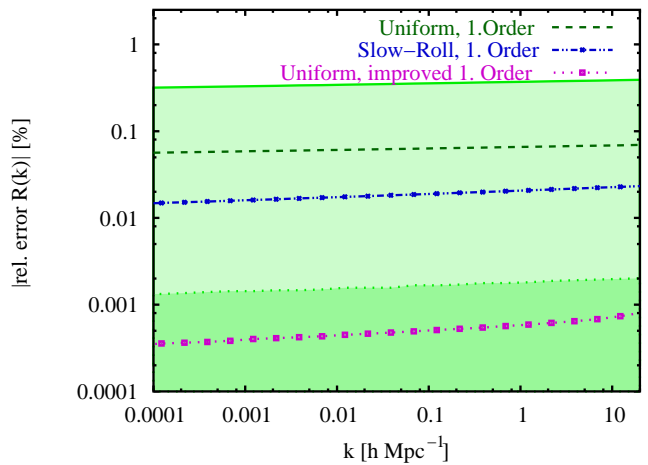}
  \caption{Relative error of the tensor to scalar ratio $R(k)$ for a
    chaotic $\frac{1}{2}m^2\phi^2$-model; parameters as specified in
    Table~\ref{tab2}. Here the dotted purple line denotes the
    (second-order) improved leading order of the uniform approximation;
    the green error band is the error estimate for the first order
    uniform approximation, the darker green band is the error
    estimate for the improved leading order result.}
  \label{fig:relerror-Rk-m20.01}
\end{figure}

The simplest chaotic model is that for a free scalar field with mass
$m$.  In this model, normalizing the amplitude of the scalar power
spectrum to the WMAP fit, as described in Appendix~\ref{units}, is
equivalent to fixing $m^2=(1.89\pm 0.21)\times 10^{-12}/8\pi G$.  The
initial value for the inflaton field is given by
$\phi(0)=16.8/\sqrt{8\pi G}$ with a small initial velocity of
$\dot{\phi}(0)=-0.1/\sqrt{8\pi G}$.  These parameters ensure that the
inflationary phase lasts long enough to provide a realistic model,
i.e., it leads to 57.655~e-folds after the $k=0.0495\mathrm{Mpc^{-1}}$
mode crosses the Hubble length.

First, we analyze the results for the power spectra $P_S(k)$ and
$P_T(k)$, plotted in Figs.~\ref{fig:PkS-m20.01}a and
\ref{fig:PkS-m20.01}b.  The exact numerical results for the power
spectra as defined in Eqns.~(\ref{PS}) and (\ref{PT}) are shown in red
(solid line), the results from the leading order uniform approximation
given in Eqns.~(\ref{PSP1}) and (\ref{PTP1}) are shown in green
(dashed line), the first order slow-roll results as defined in
Eqns.~(\ref{PSsr}) and (\ref{PTsr}) are shown in blue (dashed-dotted
line), and the improved uniform approximation results [to second-order
in the sense of Eqn.~(\ref{eq:PSimproved}) for the scalar
perturbations] are shown in purple (dotted line).  The light green
band displays our estimate of the error bound for the leading order
uniform approximation as given in Eqn.~(\ref{eq:errorkdep}).  The
error estimate for the improved uniform approximation in leading order
[see Eqn.~(\ref{errorthird})] is so close to the result itself that
the error band is not visible in this plot.  We will use these color
and linestyle assignments for the remainder of the paper.

\begin{table*}[htbp]
\caption{\label{tab3}Determination of the characteristic quantities for the 
$\frac{1}{4}\lambda\phi^4$ 
chaotic inflation model
at $k_*=0.06875~h\mathrm{\, Mpc^{-1}}=0.0495 \pm 0.0034\mathrm{\ Mpc^{-1}}$ 
($h=0.72\pm 0.05$; the  WMAP pivot scale is at $0.05\mathrm{\ Mpc^{-1}}$, 
see Ref~\cite{spergel}) 
in the various approximations; parameters: 
$\lambda=(1.75\pm 0.19)\times 10^{-13}$, 
$\phi(0)=24/\sqrt{8\pi G}$, 
$\dot\phi(0)=-1/\sqrt{8\pi G}/\mathrm{s}$, leading to 60.579 
e-folds of inflation horizon crossing of $k$=0.05/Mpc.}
\begin{ruledtabular}
\begin{tabular}{lccccc}
Approximation         
& $R(k_*)$ & $n_S(k_*)$ & $\alpha_S(k_*)$ &$n_T(k_*)$ &$\alpha_T(k_*)$\\
\hline
Numerical\footnotemark[1]             
&0.25963  &0.94999 &-0.00090        &-0.03356   &-0.01705        \\
Uniform, 1. order      
&0.25948(113)  &0.94990(3)  &-0.00089 &-0.03367(2)& -0.01694 \\
Uniform, improved 1. order\footnotemark[2]      
&0.25964(0) &  0.94999(0)& -0.00089 &-0.03356(0) & -0.01703          \\
Local approx., 0. order\footnotemark[3] 
&--  &0.94942  &-0.00092        &-0.03395   &-0.01726          \\
Local approx., 1. order\footnotemark[3] 
&--  &0.94991  &-0.00090        &-0.03340   &-0.01697          \\
Uniform, Slow-Roll redux\footnotemark[4]
&--  &0.95081  &-0.00087        &-0.03368   &-0.01706          \\
Slow-Roll, 1. order\footnotemark[5]      
&0.25969    &0.95077  &-0.00087        &-0.03285   &-0.01669          \\
Slow-Roll, 2. order\footnotemark[5]     
&--  &0.95001 &-0.00089        &-0.03354    &-0.01703          \\
\end{tabular}
\end{ruledtabular}
\footnotetext[1]{mode-by-mode integration}
\footnotetext[2]{the improvement here is to second order in powers of
  $1/\nu$, Cf.~Eqn.~(\ref{eq:PSimproved})}
\footnotetext[3]{see Eqn.~(\ref{nslocal}) and (\ref{ntlocal})}
\footnotetext[4]{see Eqn.~(\ref{nssrtp}) and Eqn.~(\ref{ntsrtp})}
\footnotetext[5]{see Eqn.~(\ref{nssr}) and (\ref{ntsr})}
\end{table*}

The exact numerical results, the improved leading order uniform
approximation, and the slow-roll approximation are almost
indistinguishable by eye in Fig.~\ref{fig:PkS-m20.01}.  The leading
order uniform approximation deviates in the amplitude from the exact
numerical result by $\sim 10\%$ as expected. The improvement strategy
performs just as predicted by Eqn.~(\ref{eq:PSimproved}): the
second-order correction reduces the error to $\sim 0.1\%$ while the
fourth-order reduces it further to $\sim 0.01\%$, for both scalar and
tensor perturbations. This consistently good behavior is due to the
fact that $\nu$ is varying slowly. Note that the improvement
strategy should be carried out to roughly match the error estimate
given by Eqn.~(\ref{errorthird}) -- fourth-order in this instance;
beyond this point the error is dominated by other contributions.

Next we investigate the scalar and tensor spectral indices as
functions of $k$.  The results are displayed in
Figs.~\ref{fig:nkS-nkT-m20.01}a and \ref{fig:nkS-nkT-m20.01}b.  The
shaded band represents the error estimate for the leading order of the
uniform approximation [see Eqn.~(\ref{errorn}), calculated with the
estimate in Eqn.~(\ref{epsbar})].  The error estimate for the scalar
spectral index is of the order $\sim 0.002\%$, therefore of the same
order as the discrepancy to the numerical result.  Note that the
improved scalar and tensor spectral index calculations, following the
discussion in Section~\ref{specint}, agree completely with the
numerical results in Table~\ref{tab2}.  The deviation of the slow-roll
approximation for the scalar spectral index in first order from the
exact numerical result is also very small, roughly $\sim 0.02\%$.  As
stated before this is not surprising: the slow-roll approximation is
expected to work well for this type of model, where the slow-roll
parameters $\epsilon$ and $\delta_1$ are very small and almost
constant.  However, the tensor spectral index has an error of more
than $1\%$.

In Fig.~\ref{fig:Rk-m20.01} the ratio $R(k)$ of tensor to scalar
perturbations as defined in Eqn.~(\ref{ratio}) is shown.  The
corresponding relative errors of the uniform approximation, of the
slow-roll approximation, and of the improved uniform approximation are
presented in Fig.~\ref{fig:relerror-Rk-m20.01}.  In
Fig.~\ref{fig:Rk-m20.01} the lines representing the different
approximations are practically indistinguishable by eye.  In contrast
to the relative errors in the power spectra ($\sim 10\%$ without
improvement and $\sim 0.01\%$ with fourth-order improvement), the
relative error in the tensor to scalar ratio $R(k)$ is already smaller
than $\sim 0.07\%$ (dashed line in Fig.~\ref{fig:relerror-Rk-m20.01})
without the improvement, and smaller than $\sim 0.001\%$, i.e.,
practically exact, with second-order improvement (solid line in
Fig.~\ref{fig:relerror-Rk-m20.01}) (Cf. Section~\ref{improve}). There
is no point in improving further since the result has already reached
the estimated error threshold (\ref{errorthird}) beyond which
improvement becomes incomplete. This behavior is a general feature of
the approximation when the inflation model leads to a slowly varying
$\nu(\eta)$.

Finally, we have listed the ratio of tensor to scalar perturbations,
the scalar and spectral spectral index, and their running in the
various approximations or simplified approximations (introduced in
Section~\ref{specind}), in Table~\ref{tab2}.  All quantities are again
evaluated at $k_*=0.0495\mathrm{\ Mpc^{-1}}$.  Error estimates for the
uniform approximation are indicated by the numbers in brackets.  For
chaotic inflation the running of the spectral indices, $\alpha_S(k)$
and $\alpha_T(k)$ is non-zero.  Note that the running of the tensor
spectral index is roughly an order of magnitude bigger than the
running of the scalar spectral index.

\subsubsection{Quartic Potential: $V(\phi)=\lambda\phi^4/4$}

\begin{figure*}[t]
  \centering
  \vspace{0.5cm}
\hspace{1cm}(a)\hspace{-1cm}
\includegraphics[width=0.97\columnwidth]{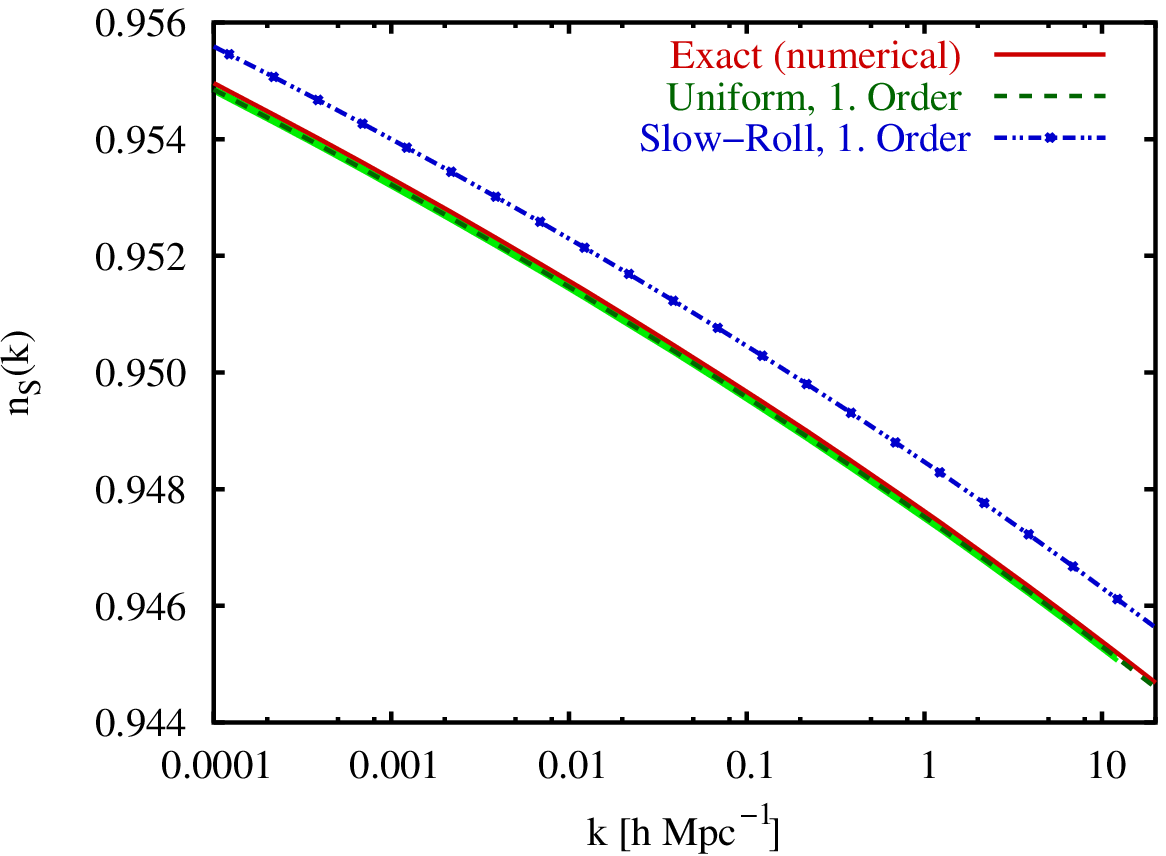}
\hspace{1cm}(b)\hspace{-1cm}
\includegraphics[width=0.97\columnwidth]{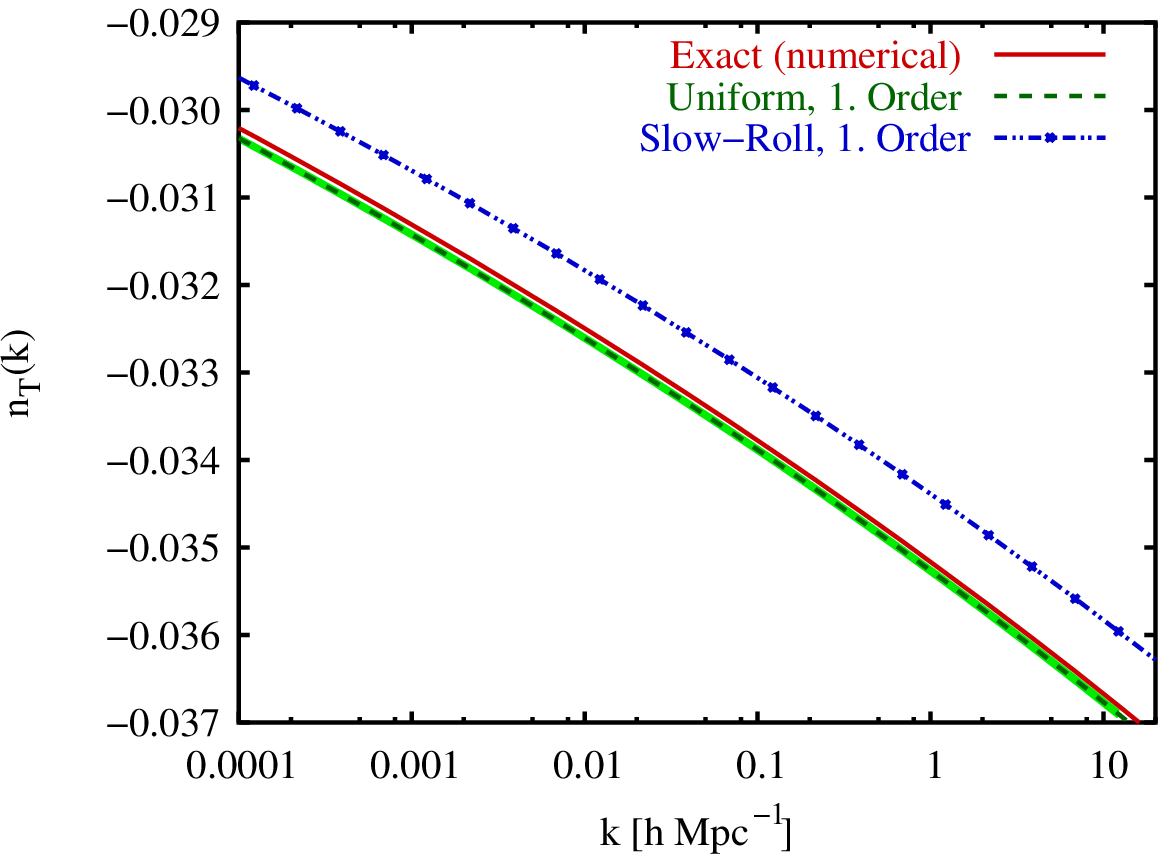}
  \caption{(a) Scalar spectral index $n_S(k)$ and 
    (b) tensor spectral index $n_T(k)$ for the quartic potential
    potential $\lambda\phi^4$, parameters as specified in
    Table~\ref{tab3}.  Solid red line: exact numerical
    results, dashed green line: uniform approximation, dashed-dotted
    blue line: slow-roll; the green band is the error estimate for
    the uniform approximation.}
  \label{fig:nk-phi4}
\end{figure*}

In the quartic model -- relative to the quadratic potential -- higher
derivatives of the potential exist. Consequently, the slow-roll
results in this case are expected to have a bigger error. In
comparison to the quadratic case, for the model considered below, the
errors are worse for the scalar spectral index ($\sim 0.02\%$ versus
$\sim 0.1\%$) and comparable for the case of the tensor spectral index
($\sim 2\%$). In contrast, the uniform approximation still provides an
accuracy of a fraction of a percent.  Although the slow-roll expansion
for this model can be improved to second-order with notably better
results, this behavior exhibits the general tendency of the slow-roll
expansion if terms arising from higher order derivatives of the
potential are significant. In terms of observational viability, the
$\phi^4$-model is under strong pressure from combined analysis of the
WMAP CMBR anisotropy data and data from galaxy clustering (see e.g.,
Refs.~\cite{Peiris,LeachLiddle,Tegmark:2004qd}).

The results for the spectral indices are displayed in
Figs.~\ref{fig:nk-phi4}a and \ref{fig:nk-phi4}b.  As in the previous
example the leading-order uniform approximation is very close to the
exact numerical results.  However, the first-order slow-roll result
does not match as closely as for the quadratic potential.  As done
earlier for the $\phi^2$-model, we have listed the various
characteristic quantities in Table~\ref{tab3}.

\subsection{Inflationary Model with a $C^2$-Potential Function}

As the last example we investigate a toy model with a continuous
potential function with continuous first and second derivatives and a
jump in the third derivative at a specific value of $\phi$.  Generally
speaking, dynamical changes in the potential of the field driving
inflation can be induced by couplings to other degrees of freedom. For
example, in hybrid models a phase transition is used to terminate
inflation. If a dynamical transition happens at cosmologically
relevant scales, i.e., well before the end of inflation, it leaves a
clear signature in the power spectra and spectral indices.  Such a
transition may be naturally realized in multi-field models of
inflation (see e.g., Ref.~\cite{Hunt:2004vt}).  Other examples are
steps in the potential (see e.g., Ref.~\cite{wms,Adams:2001vc}),
leading to oscillations in the primordial power spectra and spectral
indices.  In such models $\nu^2$ cannot be considered as a constant,
but can display sudden changes.

\begin{figure}[b]
  \centering
  \includegraphics[width=0.97\columnwidth]{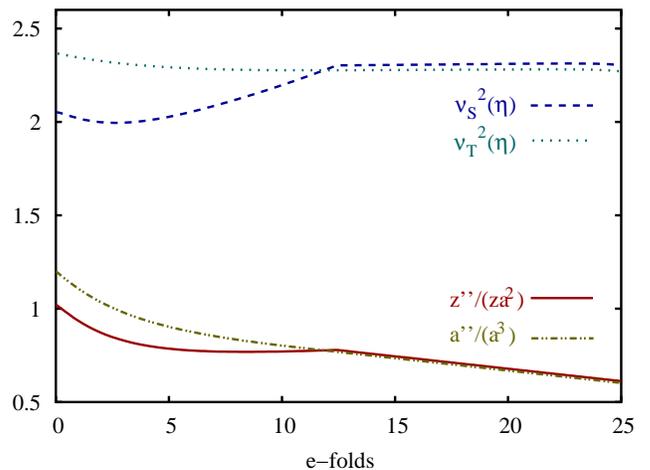}
  \caption{$z''/(za^2)$ and $\nu^2_S$ and $a''/a^3$ and $\nu_T$ for 
    the $C^2$-potential; the point $\phi_*$ is reached at $N\approx
    12.4$. The beginning in time of the numerical calculation is at
    $N=0$. The inflationary attractor is reached at $N\simeq 1.8$,
    checked by varying $\dot\phi(0)$ and determining at which e-fold
    the $\phi$ behavior becomes independent of the initial
    velocities. The quantities $a''/a^3$ and $\nu_T$, relevant for
    tensor perturbations, are much smoother than the corresponding
    quantities $z''/za^2$ and $\nu_S$ for the scalar perturbations, 
    leading to smaller errors in the approximations as discussed in
    the text.} 
  \label{fig:zppza2-nu2}
\end{figure}

\begin{figure*}[htbp]
  \centering
  \vspace{0.5cm}
\hspace{1cm}(a)\hspace{-1cm}
\includegraphics[width=0.97\columnwidth]{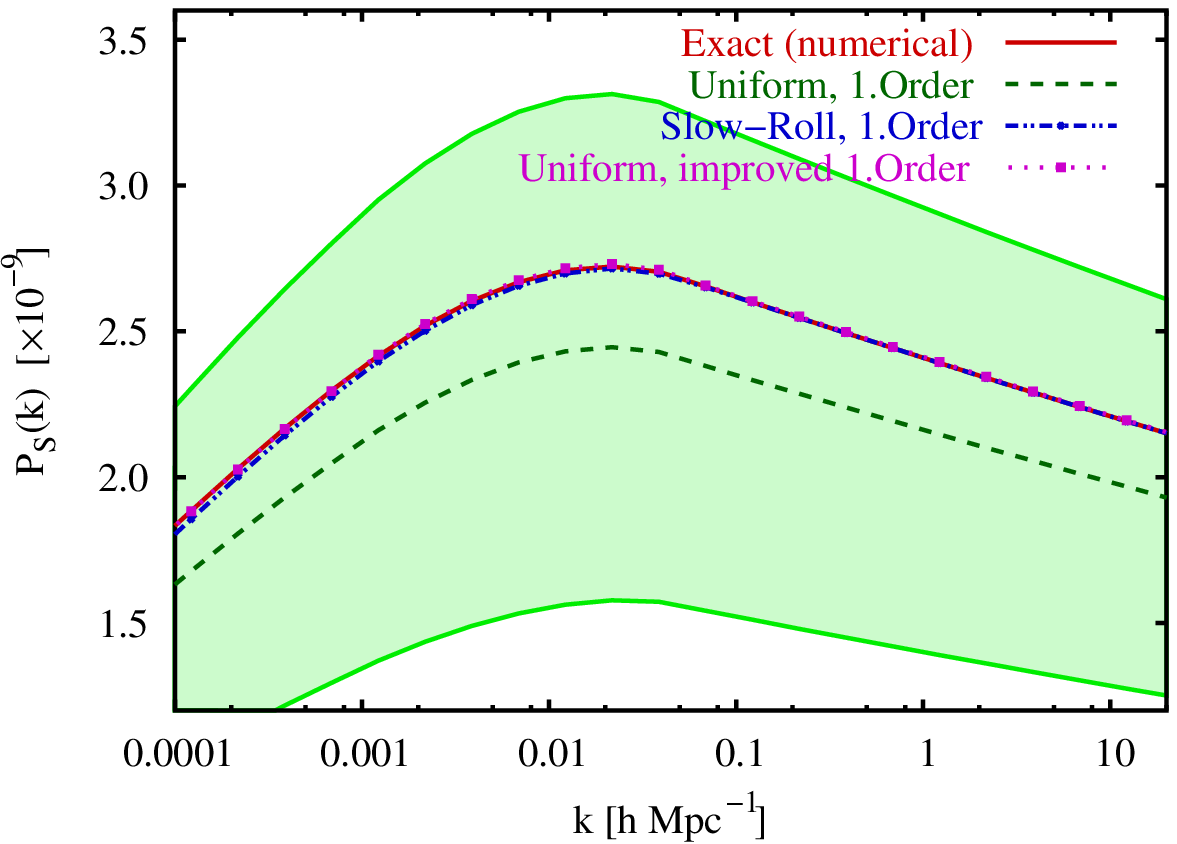}
\hspace{1cm}(b)\hspace{-1cm}
\includegraphics[width=0.97\columnwidth]{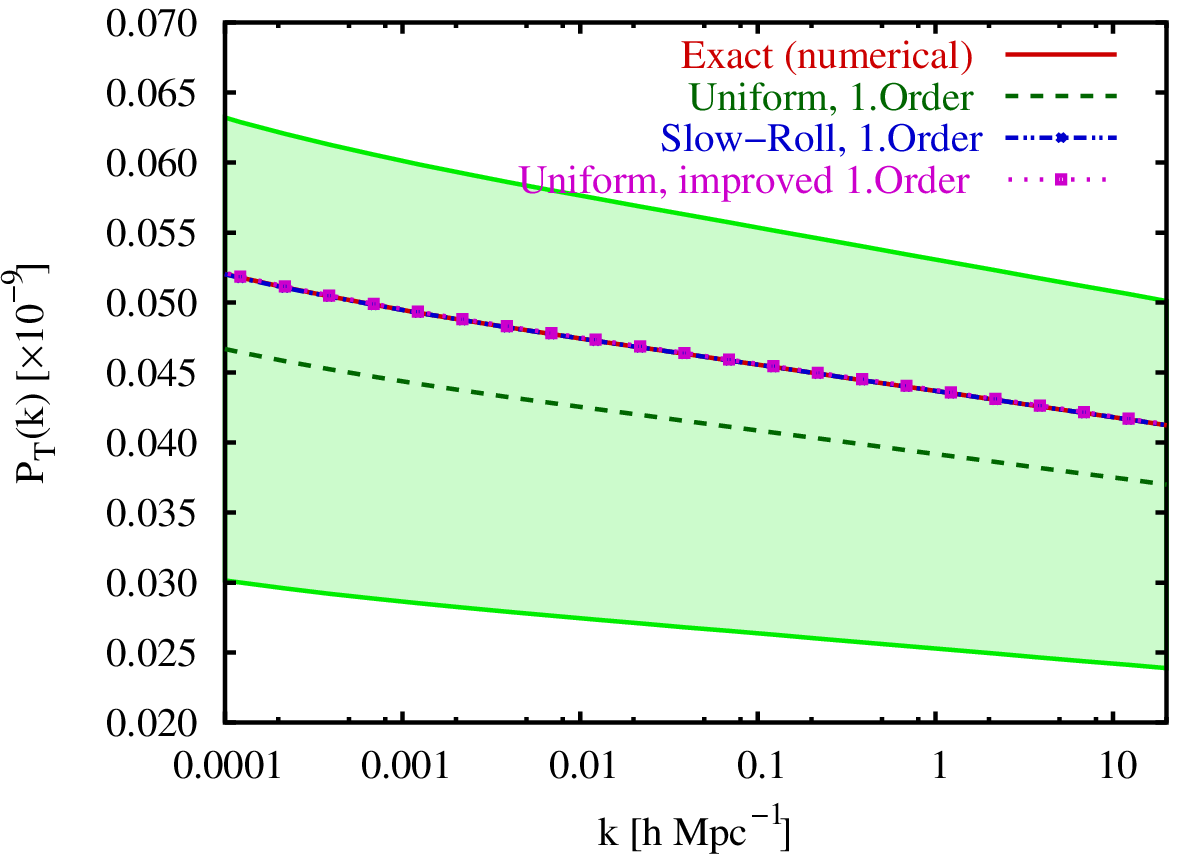}
    \caption{(a) Scalar power spectrum $P_S(k)$ and (b) tensor power
    spectrum $P_T(k)$ for the $C^2$-potential in Eqns.~(\ref{eq:Vgtr})
    and (\ref{eq:Vsmr}); parameters: $\alpha=-100$, $m^2=(1.90\pm
    0.21)\times 10^{-12}/8\pi G$, $\phi_*=15.2/\sqrt{8\pi G}$,
    $\phi(0)=17.5/\sqrt{8\pi G}$, $\dot{\phi}(0)=-0.2/\sqrt{8\pi
    G}\mathrm{s}$. Solid red line: exact numerical results, dashed
    green line: uniform approximation, dashed-dotted blue line:
    slow-roll; the green band is the estimate for the error bound for
    the (unimproved) uniform approximation.  Again, the exact results
    and the results from the improved uniform and slow-roll
    approximation are on top of each other.}
  \label{fig:Pk-glued}
\end{figure*}

\begin{figure*}[htbp]
  \centering
  \vspace{0.5cm}
\hspace{1cm}(a)\hspace{-1cm}
\includegraphics[width=0.97\columnwidth]{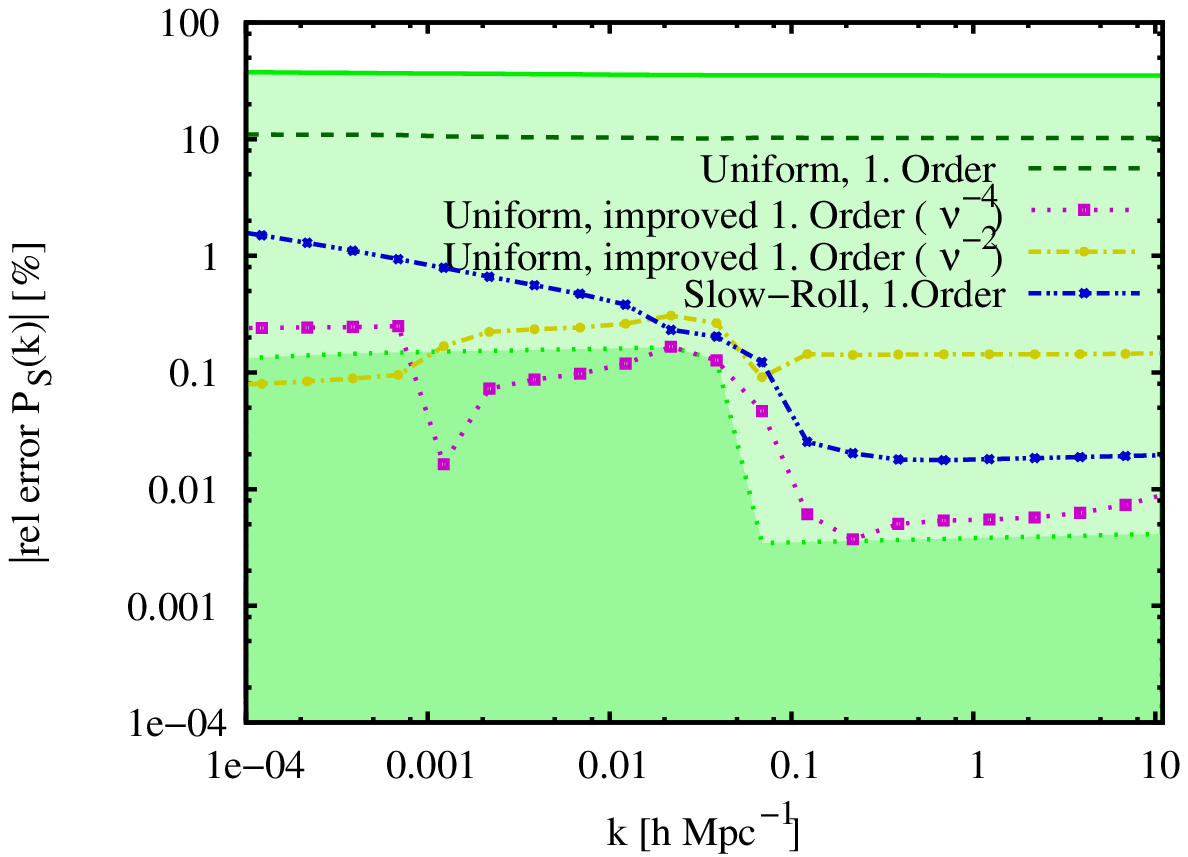}
\hspace{1cm}(b)\hspace{-1cm}
\includegraphics[width=0.97\columnwidth]{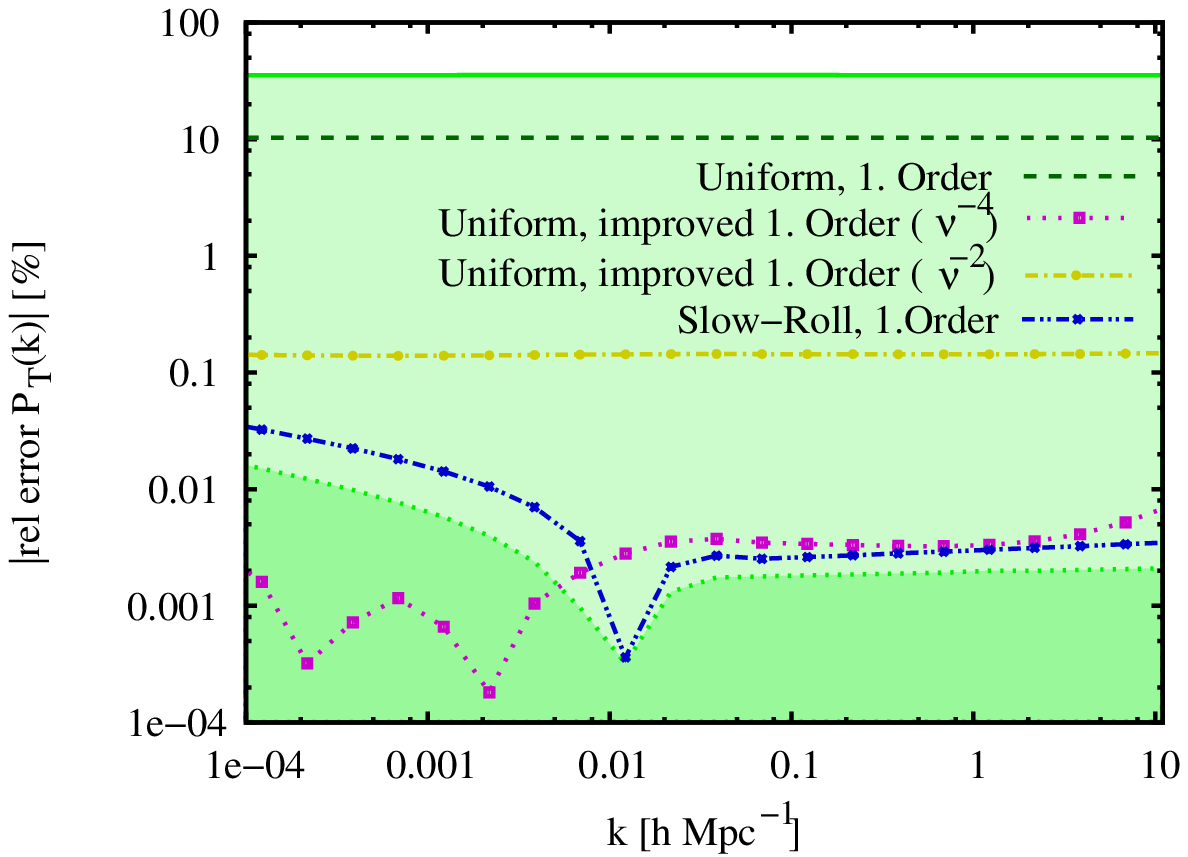}
  \caption{(a) Relative errors for the scalar power spectrum $P_S(k)$
    and (b) tensor power spectrum $P_T(k)$ for the $C^2$-potential,
    Eqns.~(\ref{eq:Vgtr}) and (\ref{eq:Vsmr}). In both cases, the
    light green band denotes the estimated error bound for the
    first-order uniform approximation (\ref{eq:errorkdep}) and the
    dark green band denotes the best estimated error for the
    improvement procedure (\ref{eq:PSimproved}). The results are
    nicely consistent with these estimates showing where the
    second-order improvement can be enhanced by going to higher order,
    and where it cannot.}
  \label{errsc}
\end{figure*}

Rather than taking one of the potentials mentioned above we consider
here a toy potential that is smoother in the sense that oscillations
in $\nu^2$ or $z''/z$ are avoided:
\begin{eqnarray}
V_>(\phi)&=&\frac{1}{4}m^2\phi_*^2(\alpha-1)
+\frac{2}{3}m^2\phi_*(1-\alpha)\phi
\nonumber \\
&&+\frac{1}{2}\alpha m^2 \phi^2 
+\frac{1}{12\phi^2_*}m^2(1-\alpha)\phi^4, \label{eq:Vgtr}\\
V_<(\phi)&=&\frac{1}{2}m^2\phi^2 \label{eq:Vsmr},
\end{eqnarray}
where $V(\phi)=V_>(\phi)$ for $\phi>\phi_*$ and $V(\phi)=V_<(\phi)$
for $\phi<\phi_*$. The potential is constructed in such a way that
$V_>(\phi_*)=V_<(\phi_*)$, $V'_>(\phi_*)=V'_<(\phi_*)$ and
$V''_>(\phi_*)=V''_<(\phi_*)$ but $V'''_>(\phi_*)\neq V'''_<(\phi_*)$.
Thus there is a finite jump in the third derivative of the potential.

We present numerical results with parameters chosen specifically to
demonstrate the general effect of a more rapidly changing $\nu$.
During the evolution in this potential the parameters $\epsilon$ and
$\delta_i$ are not constant (not even approximately); $\delta_1$
cannot be considered small when the inflaton field fulfills
$\phi>\phi_*$ (with the parameters below, $|\delta_1|$ can be as large
as $0.14$).

The parameters chosen to specify the model are: $\alpha=-100$,
$m^2=(1.90\pm 0.21)\times 10^{-12}/8\pi G$,
$\phi_*=15.2/\sqrt{8\pi G}$,
$\phi(0)=17.5/\sqrt{8\pi G}$,
$\dot{\phi}(0)=-0.2/\sqrt{8\pi G}/\mathrm{s}$. With this
choice of parameters the number of e-folds counted from $k_*=0.0495\ 
\mathrm{Mpc}^{-1}$ is $57.320$.

\begin{figure*}[htbp]
  \centering
  \vspace{0.5cm}
\hspace{1cm}(a)\hspace{-1cm}
\includegraphics[width=0.97\columnwidth]{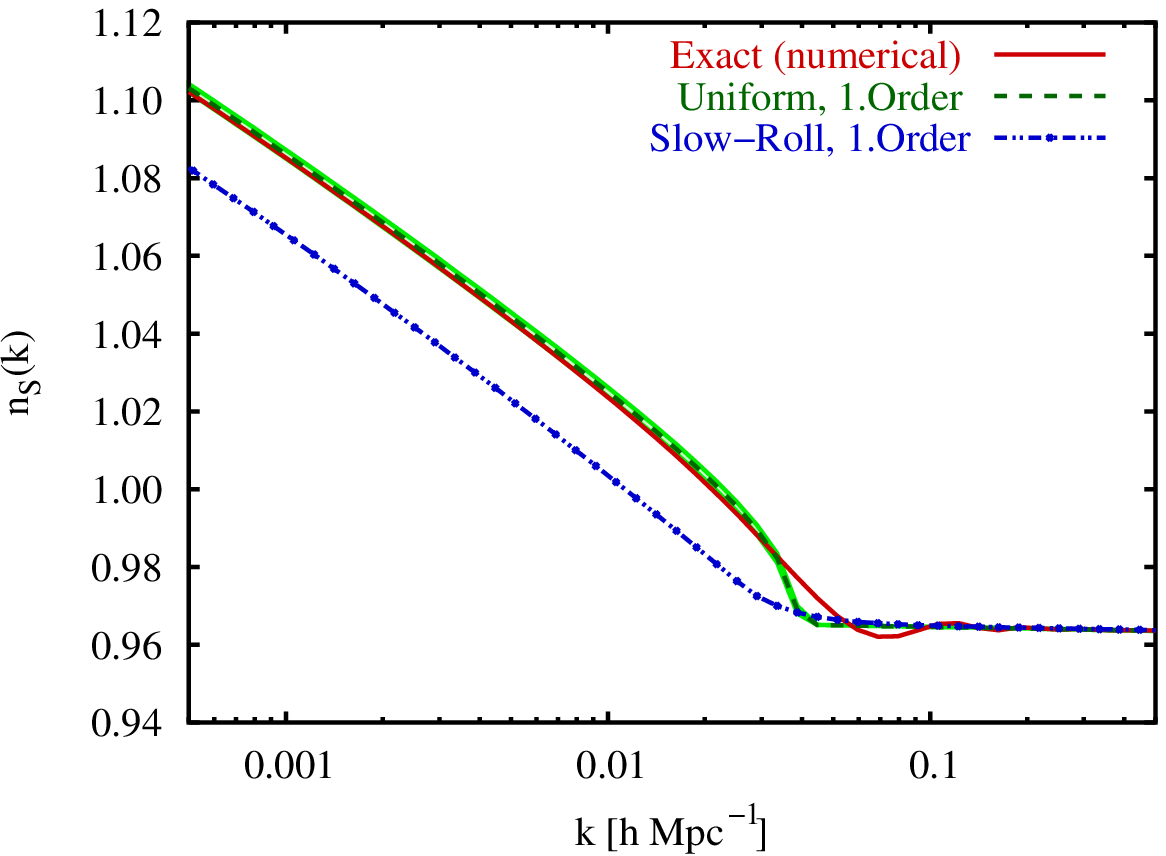}
\hspace{1cm}(b)\hspace{-1cm}
\includegraphics[width=0.97\columnwidth]{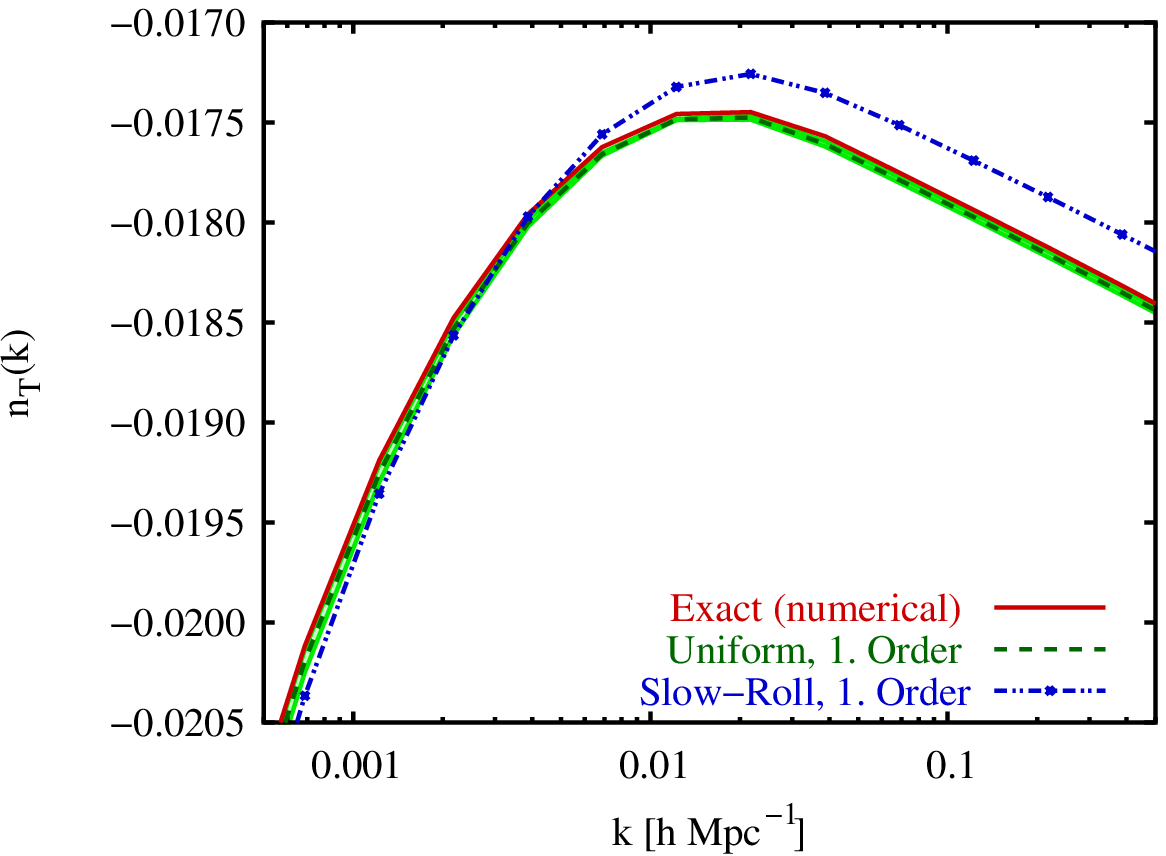}
  \caption{(a) Scalar spectral index $n_S(k)$ and 
    (b) tensor spectral index $n_T(k)$ for the $C^2$-potential in
    Eqns.~(\ref{eq:Vgtr}) and (\ref{eq:Vsmr}) in the region around the
    kink; parameters specified as in Fig.~\ref{fig:Pk-glued}.  Solid
    red line: exact numerical results, dashed green line: uniform
    approximation, dashed-dotted blue line: slow-roll; the green band
    is the error estimate for the uniform approximation.}
  \label{fig:nk-glued}
\end{figure*}

The relevant time-dependent terms $\nu_S^2$ and $z''/(za^2)$ in the
scalar and $\nu_T^2$ and $a''/a^3$ in the tensor mode equations are
displayed in Fig.~\ref{fig:zppza2-nu2} as a function of expansion
e-folds ($z''/z$ and $a''/a$ have been divided out by $a^2$ to filter
out the exponential growth of the scale factor). The point $\phi_*$ is
reached at $N\approx 12.4$ (note that in this plot $N=0$ defines the
beginning of the numerical calculation).  Due to the jump in the third
derivative both quantities for the scalar perturbations display a kink
at this point. The qualitative behavior is also different on either
side of the kink. While $z''/z$ has a kink, $a''/a$ is completely
well-behaved.  Note that $z''/z$, as e.g., expressed in
Eqn.~(\ref{zdpz}) as an exact expression in terms of the slow-roll
parameters, is more sensitive to higher derivatives of the potential
than $a''/a$ [Cf. Eqn.~(\ref{adpa})].  Thus we can expect the effects
of the change in the potential at $\phi=\phi_*$ to be amplified in the
scalar power spectrum relative to the tensor power spectrum.

The results for the scalar and tensor power spectrum are displayed in
Figs.~\ref{fig:Pk-glued}a and \ref{fig:Pk-glued}b.  With the same
conventions as in Fig.~\ref{fig:PkS-m20.01}, the different
approximations (leading and improved leading order of the uniform
approximation and the slow-roll approximation) are compared to the
exact numerical results.  The scalar power spectrum (see
Fig.~\ref{fig:Pk-glued}a) shows a significant deviation from a
power-law shape.  Up to $k\approx 0.025\ h\mathrm{Mpc}^{-1}$ the
spectrum rises, reaches a maximum and falls off for larger $k$.  As in
the previous examples, the leading order of the uniform approximation
has an amplitude error of roughly $10\%$ with respect to the exact
numerical results.  The (second-order) improved leading order uniform
approximation, however, lies almost on top of the numerical results. 
Remarkably, although the shape deviates from a simple power-law
behavior quite significantly, the improvement strategy is still
effective.

The relative errors are shown in Figs.~\ref{errsc}a and \ref{errsc}b.
The error behavior divides into two regimes, to the left and the right
of $k\sim 0.05$~$h$Mpc$^{-1}$.  The behavior to the right is that of a
$\phi^2$-model [Cf.  Eqn.~(\ref{eq:Vgtr})] while the behavior to the
left is that of a polynomial potential with linear, quadratic, and
quartic terms (\ref{eq:Vsmr}).  Note that the error estimate from
Eqn.~(\ref{errorthird}) changes sharply across this divide, by more
than an order of magnitude, from $\sim 0.1\%$ to $\sim 0.005\%$.  To
the left, this error estimate shows that there is no point in
attempting a correction beyond second-order using
Eqn.~(\ref{eq:PSimproved}), consistent with the results shown for
second and fourth-order corrected spectra.  To the right, the
smallness of the error estimate is consistent with the improved
quality of the fourth-order results.  Results for the tensor spectrum
are qualitatively similar.  As expected, the uniform approximation
improves on the slow-roll result to the left of $k\sim
0.05$~$h$Mpc$^{-1}$, since the slow-roll assumptions are violated in
this region.

\begin{figure*}[t]
  \centering
  \vspace{0.5cm}
\hspace{1cm}(a)\hspace{-1cm}
\includegraphics[width=0.97\columnwidth]{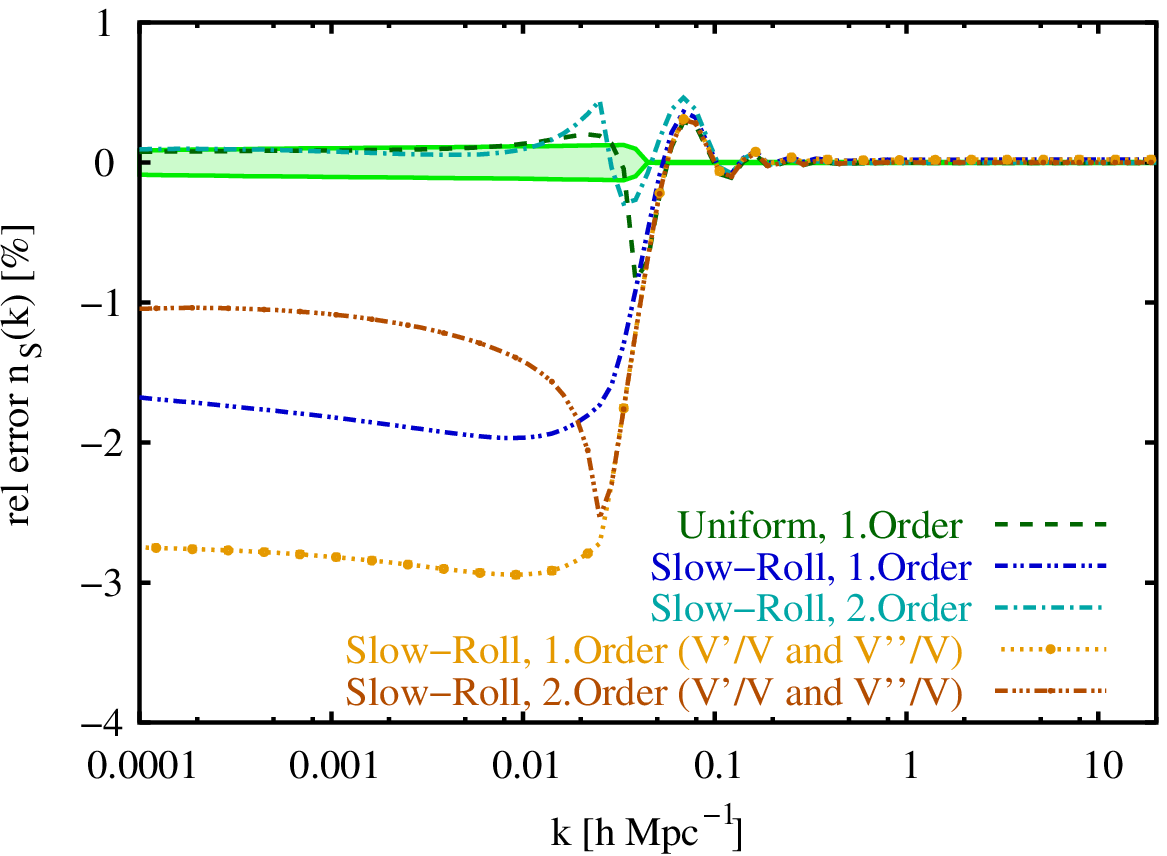}
\hspace{1cm}(b)\hspace{-1cm}
\includegraphics[width=0.97\columnwidth]{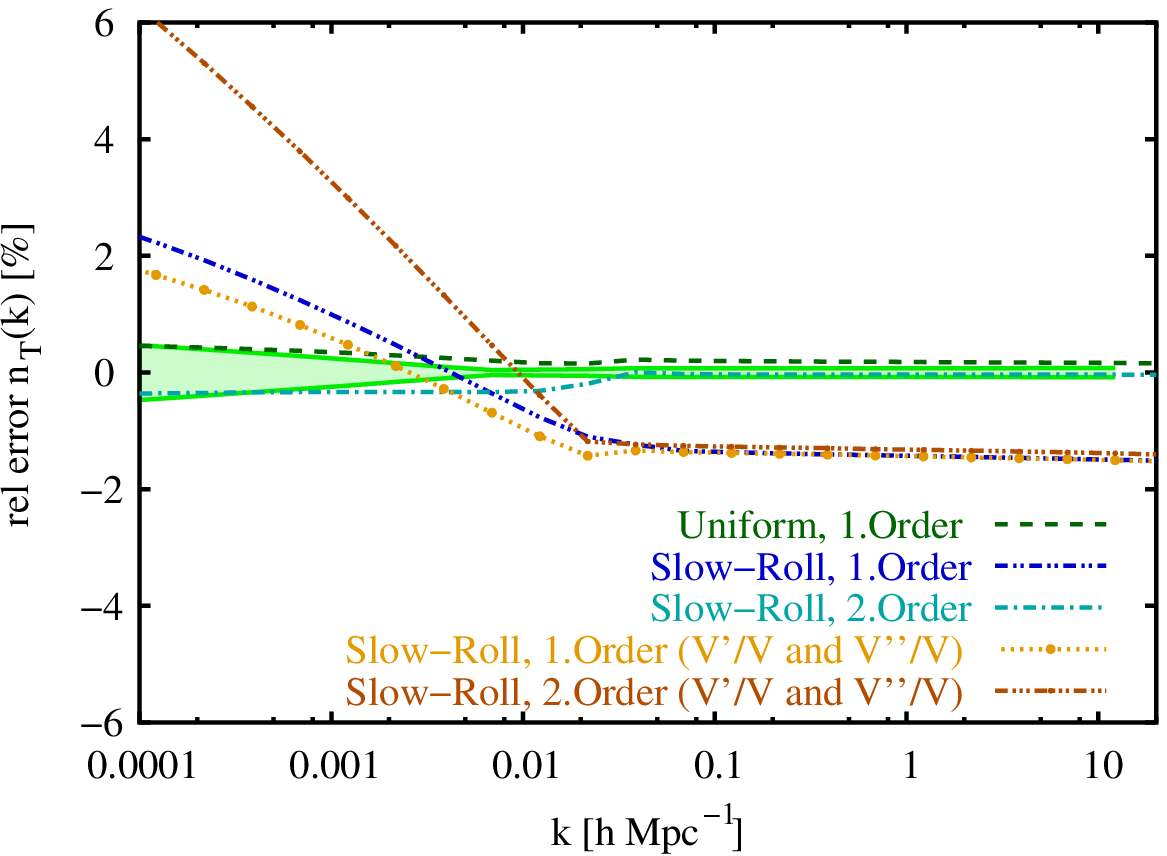}
  \caption{Relative error of the (a) scalar and (b) tensor spectral
    index (see Figs.~\ref{fig:nk-glued}a~and~\ref{fig:nk-glued}b) for
    the $C^2$-potential in Eqns.~(\ref{eq:Vgtr}) and (\ref{eq:Vsmr}).
    The green band is the error estimate for the uniform
    approximation. In addition to the uniform and the slow-roll
    approximation we also show the slow-roll approximation including
    higher order derivatives of $V(\phi)$ [see Eqns.~(\ref{epsV}) and
    (\ref{delV})]. Light brown dotted line: first order slow-roll,
    dark brown dashed-dotted line: second order slow-roll.}
  \label{fig:relerrornk-glued}
\end{figure*}

The spectral indices are displayed in Figs.~\ref{fig:nk-glued}a and
\ref{fig:nk-glued}b.  The potential $z''/z$ (see
Fig.~\ref{fig:zppza2-nu2}) leads to a blue scalar spectrum for smaller
momenta and a red scalar spectrum for larger momenta
(Fig.~\ref{fig:nk-glued}a).  For the spectral index the uniform
approximation in leading order is remarkably close to the exact
numerical result, more or less independent of $k$.  It only deviates
slightly at $k\approx 0.04\,h\,\mathrm{Mpc}^{-1}$ for the scalar
spectral index.  In Fig.~\ref{fig:relerrornk-glued} the relative
errors of the uniform approximation and the slow-roll approximation
are displayed.  Away from the transition $k$-value, the relative error
is smaller than $\sim 0.2\%$ for the scalar spectral index and smaller
than $\sim 0.5\%$ for the tensor spectral index.  The slow-roll
approximation, by comparison, deviates by $\sim 2\%$ from the exact
numerical results.

\begin{figure}[t]
  \centering
  \vspace{0.5cm}
\includegraphics[width=0.97\columnwidth]{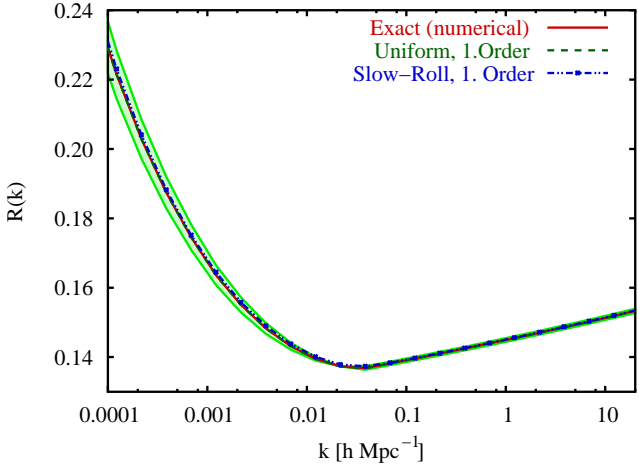}
  \caption{Ratio $R(k)$ of tensor to scalar perturbations 
    for the $C^2$-potential in Eqns.~(\ref{eq:Vgtr}) and (\ref{eq:Vsmr}); the
    green band is the error estimate for the uniform approximation to
    leading order; the relative difference between all three
    approximations is below $2\%$ (see
    Fig.~\ref{fig:rel-error-Rk-glued}), making it difficult to
    distinguish the curves. }
  \label{fig:Rk-glued}
\end{figure}

\begin{figure}[t]
  \centering
  \vspace{0.5cm}
\includegraphics[width=0.97\columnwidth]{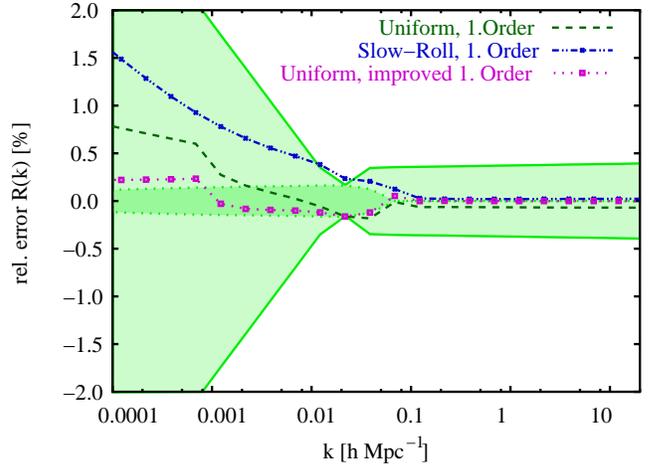}
    \caption{Relative error for the ratio $R(k)$ (ratio of tensor to
    scalar perturbations for the $C^2$-potential in
    Eqns.~(\ref{eq:Vgtr}) and (\ref{eq:Vsmr}); the green band is the
    error estimate for the uniform approximation to leading order, the
    darker green band is the estimate for the improved
    leading order.}
  \label{fig:rel-error-Rk-glued}
\end{figure}

In all the computations so far, the background equations were not
approximated to obtain $H(t)$ and $\phi(t)$ and its derivatives, but
were solved numerically. Based on a Taylor expansion in the potential
$V(\phi)$ and its derivatives, Liddle and Lyth~\cite{LL92} introduced
an approximation for the background equations.  This approach
(formalized and expanded to higher orders by Liddle, Parsons, and
Barrow~\cite{LPB}) leads to a simplification of the slow-roll
parameters in the following form:
\begin{eqnarray}\label{epsV}
\epsilon &=& \frac{1}{2}\left(\frac{V'}{V}\right)^2-\frac{1}{3}
\left(\frac{V'}{V}\right)^4+\frac{1}{3}\frac{V'^2V''}{V^3},\\
\label{delV}
\delta_1 &=& \frac{1}{2}\left(\frac{V'}{V}\right)^2-\frac{V''}{V}
-\frac{2}{3}\left(\frac{V'}{V}\right)^4\nonumber\\
&&-\frac{1}{3}
\left(\frac{V''}{V}\right)^2+\frac{4}{3}\frac{V'^2V''}{V^3}.
\end{eqnarray}
Here we have followed the conventions of Stewart and Lyth~\cite{sl}.
Being an asymptotic expansion, this approximation can sometimes lead to
extra errors. For the last model, we have calculated the spectral
indices with this additional approximation and the degradation in
relative error is shown in Fig.~\ref{fig:relerrornk-glued}.

The ratio $R(k)$ of tensor to scalar perturbations is depicted in
Fig.~\ref{fig:Rk-glued}, while the corresponding relative errors for
the different approximations are shown in
Fig.~\ref{fig:rel-error-Rk-glued}. The uniform and the slow-roll
approximation are both quite close ($\sim 1\%$ error) to the exact
numerical result, even though the variations in $R(k)$ are not small.
Following Section~\ref{improve} the accuracy of the leading-order
uniform approximation for the ratio $R(k)$ can be improved using
Eqn.~(\ref{eq:PSimproved}) and the corresponding equation for tensor
perturbations. We have not displayed this improved ratio in
Fig.~\ref{fig:Rk-glued}, since it would be almost indistinguishable
from the exact numerical result. However,
Fig.~\ref{fig:rel-error-Rk-glued} shows that the relative error of the
second-order improved leading order of the uniform approximation is
smaller than $\sim 0.3\%$ over the whole $k$-range, consistent with
the error estimate (\ref{errorthird}).

\subsection{Consistency relations}

Single field inflation is characterized by degeneracies in the
observable parameters, such as the tensor to scalar ratio $R(k)$, the
scalar spectral index $n_S$, the tensor spectral index $n_T$, and their
respective running.  Not all the parameters are independent; they are
connected by so-called ``consistency relations.''  For power-law
inflation, e.g., the tensor spectral index is related to the scalar
spectral index via the special result $n_T=n_S-1$.  In the context of
slow-roll inflation, consistency relations between, e.g., $R(k)$ and
$n_T(k)$ have been derived (see e.g., Ref.~\cite{LL92} for an early
derivation, note, however, that Liddle and Lyth define $R(k)$ as the
ratio of the quadrupole moments.  Since this definition introduces
further uncertainties through the transfer functions here we use the
amplitudes of the power spectra themselves).  A more general
expectation beyond slow-roll is that $R(k)$ is some function of
$n_S(k)$ and $n_T(k)$.

\begin{figure}[tbhp]
  \centering
  \includegraphics[width=0.97\columnwidth]{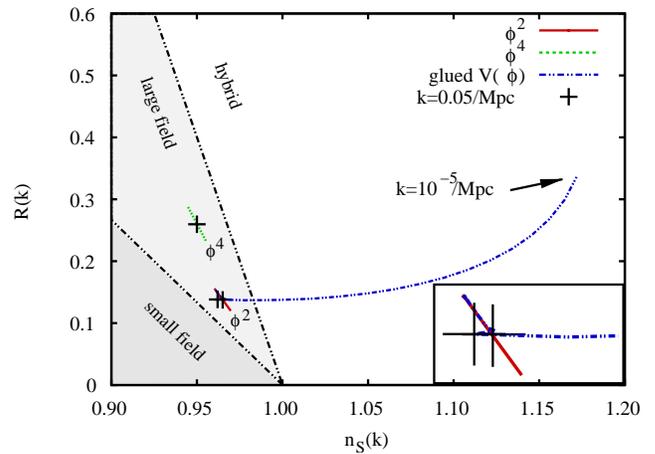}
  \caption{\label{fig:R-nS}
    Tensor to scalar ratio $R(k)$ against the scalar spectral index
    $n_S(k)$ for the exact numerical results of the inflation models
    investigated here; via the $k$ dependence (running of the spectral
    index and the tensor to scalar ratio) each model produces a line --
    the solid red line denotes the $\phi^2$-model, the dashed green
    line the $\phi^4$-model and the dashed-dotted blue line the glued
    $C^2$-potential; the black crosses denote the central WMAP pivot
    $k=0.05\ \mathrm{Mpc}^{-1}$.  The two dotted lines delimit small
    field, large field and hybrid inflation models according to the
    classification in Ref.~\cite{Dodelson:1997hr}.  In the insert we
    show a zoom into the region where the $C^2$-potential goes over
    into the $\phi^2$-potential.  It is interesting to note that at
    this point the curve bends over sharply.}
\end{figure}

\begin{figure}[tbph]
  \centering
  \includegraphics[width=0.97\columnwidth]{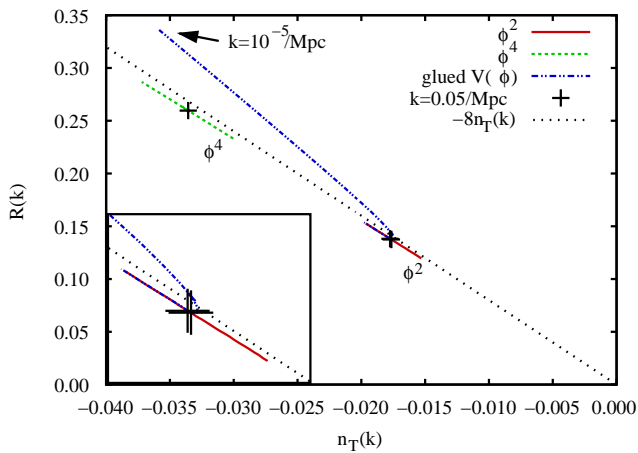}
\caption{\label{fig:R-nT}Tensor to scalar ratio $R(k)$ against the
  tensor spectral index $n_T(k)$; specifications as in
  Fig.~\ref{fig:R-nS} except that the dotted line here denotes $R=
  -8n_T$. By fitting the $\phi^2$- and the $\phi^4$-model results,
  they are found to lie on a straight line $R=-(7.757\pm 0.004)n_T$.
  Although the numerical prefactor is different from the slow-roll
  estimate of $-8$, the degeneracy is still obvious. As in the
  previous figure, the insert shows a zoom into the region where the
  curve of the $C^2$-potential bends over to join the line from the
  $\phi^2$-potential.}  
\end{figure}

Following Ref.~\cite{LLKCBA} we introduce rescaled amplitudes via
\begin{eqnarray}
  A_S(k)&=&\frac{2}{5}P^{1/2}_\mathcal{R}(k),\\
  A_T(k)&=&\frac{1}{10}P^{1/2}_h(k), 
\end{eqnarray}
leading, in the first order slow-roll expansion, to a consistency
relation between $R(k)$ and $n_T(k)$ of the form:
\begin{equation}
  n_T\simeq-2\frac{A_T^2}{A_S^2}.
\end{equation}
Or, adapted to our notation,
\begin{equation}
R(k)=\frac{P_h(k)}{P_\mathcal{R}(k)}=\frac{8P_T(k)}{P_S(k)}\simeq 
-8n_T(k).  
\label{eq:consiRnT}
\end{equation}
The equality in this relation is only valid when the first-order
slow-roll quantities are used, even for power-law inflation. Note also
that the relation is $k$-dependent in general. 

Inspired by the idea of consistency relations between measurable
quantities from the microwave background, Dodelson et
al.~\cite{Dodelson:1997hr} classified inflationary models depending on
the relative magnitudes of the spectral indices $n_S$ and $n_T$, and
the ratio of the amplitudes of the power spectra $R(k)$. In terms of
slow-roll parameters, the classification is based on the relative
magnitudes of $\epsilon$ and $\delta_1$. Using
Eqn.~(\ref{eq:consiRnT}), relations between slow-roll parameters can
be translated to inequalities between $R(k)$ and $n_S$.  The three
model types are then characterized by:

\begin{enumerate}
\item Small field models: 

$-\delta_1<-\epsilon$ or $R\lesssim -\frac{8}{3}(n_S-1)$, 

\item Large field models: 

$-\epsilon<\delta_1<\epsilon$ or 
$-\frac{8}{3}(n_S-1)\lesssim R \lesssim -8(n_S-1)$,

\item Hybrid models: 

$\epsilon<-\delta_1$ or $-8(n_S-1)\lesssim R$.
\end{enumerate}

In Fig.~\ref{fig:R-nS} we show $R(k)$ as a function of $n_S(k)$ for
the three models investigated in the previous sections. The
$k$-dependence, i.e., the running of the quantities, produces lines,
rather than single points in the $R-n_S$ plane.  For the single-field
chaotic models these lines are straight. This is not the case,
however, for the $C^2$-potential where the $k$-dependence is not a
straight line, but a curve.  For all three models we have marked the
location of the central pivot $k_*=0.05~\mathrm{Mpc}^{-1}$ with a
cross. In all cases the total number of e-folds is fixed to be $N\sim
60$ counted from the pivot scale.

We can also test the consistency relation $R\simeq -8n_T$ of slow-roll
inflation.  In Fig.~\ref{fig:R-nT}, $R(k)$ is plotted as a function of
$n_T(k)$.  The slow-roll dominated chaotic models are indeed very
close to the slow-roll consistency relation $R\simeq -8n_T$ (the
dotted line in Fig.~\ref{fig:R-nT}), but not the $C^2$-model. A
detailed understanding of the consistency relations and their
observational value requires further study.

Regarding observations, the allowed range in the $R$-$n_S$ plane is
already constrained by WMAP, the SDSS Galaxy clustering information
and the SDSS Ly$\alpha$ forest measurements (see e.g.,
Ref.~\cite{Tegmark:2004qd}). An observation of parametric degeneracy by
future CMBR experiments, e.g., PLANCK~\cite{PLANCK}, would provide
strong support for inflation as the source of primordial
fluctuations. 

\section{Conclusions}
\label{conclusion}
      
In this paper we have presented different approaches to the
computation of primordial power spectra and the corresponding spectral
indices and their running, with a view to understanding and controlling
the various sources of error in the calculations. In addition, the
ratio of the tensor to scalar power spectra and the related
consistency relation were investigated.

We have implemented an efficient and accurate method for exact
mode-by-mode integration utilizing results from the uniform
approximation to set up the initial conditions. The uniform
approximation, introduced for calculating inflationary perturbations
in~\cite{hhjm,hhhjm}, was numerically investigated in detail. We
showed that the leading order results in this approximation for the
power spectrum can be easily improved for well-behaved $\nu_S$ and
$\nu_T$ using previously obtained results for the case of constant
$\nu$ (the results for the spectral indices $n_S$ and $n_T$ are
already excellent and do not need to be improved). Our numerical and
semi-analytic results for the power spectra and spectral indices agree
within $0.1\%$.  Thus, the primordial power spectra can be determined
at the same level of accuracy as the transfer functions.

A key feature of the uniform approximation is the existence of an
error control theory, which is missing in the slow-roll approximation.
At leading-order in the uniform approximation, we showed how to
implement a useful approximate error bound for the power spectrum.  In
addition, we provided an error estimate for the power spectra obtained
from the improved first order uniform approximation and the spectral
indices. These error estimates are based on our knowledge of the
ultra-local corrections from the second order uniform approximation
results. Thus, these estimates are much tighter than the general
leading order error bounds from the uniform approximation.
 
We have analyzed in detail three different classes of inflationary
models: power-law models, chaotic models, and a model with a
$C^2$-potential.  We used the power-law model, for which exact
analytical results exist, to demonstrate the accuracy of our numerical
implementations of the exact mode-by-mode integration, the uniform
approximation, and the slow-roll approximation.  The deviation of the
analytic and numerical results was in general very small, around 1
part in $10^{6}$.  Two scale-free chaotic inflationary models were
chosen as representatives for common slow-roll models: For these
models the slow-roll approximation was excellent.  As a final example,
we constructed a $C^2$-potential which had two dynamical phases
patched together in a relatively smooth manner.  In the first phase,
where $\nu$ changes more rapidly than in the second phase, the uniform
approximation was much more accurate than the slow-roll approximation,
while in the second phase both approximations produced very good
results with small errors.

Up-coming high-precision CMBR measurements will provide data to
constrain the zoo of inflation models. An accurate and fast code for
calculating primordial power spectra, spectral indices, and their
running will be crucial to this analysis. We are developing an
interface to connect our code to Boltzmann solvers~\cite{cmbfast} in
order to generate the $C_l$'s directly.  In addition, we will use the
code to test the robustness of the information obtained on the
inflationary equation of state from the measured power spectrum by our
recently introduced non-parametric reconstruction program~\cite{hhj}.
These two complementary approaches provide new precision probes of the
first moments of our Universe.

\section{Acknowledgments}

The authors thank Scott Dodelson for useful discussions and
encouragement to write a code for exact mode-by-mode integration.  We
are grateful to Max Tegmark for permission to reproduce
Fig.~\ref{obsfig}.  AH gratefully acknowledges the warm hospitality
and stimulating atmosphere at the Los Alamos National Laboratory
(LANL). AH thanks J\"urgen Baacke for helpful discussions and
continuous encouragement.  The work of AH at LANL has been supported
by a DAAD short-term scholarship for graduate students and by the
University of Dortmund. This research is supported by the Department
of Energy, under contract W-7405-ENG-36.

\begin{appendix}

\section{Numerical Implementation Details}
\label{app}

\subsection{Comment on the Momentum Discretization}    
\label{mom}
The actual momentum discretization chosen for numerical work is
arbitrary but it is a good idea to adjust the chosen values of the
momenta so that the relation
\begin{equation}
\bar{\nu}^2(\bar{\eta})=k^2\bar{\eta}^2 \Rightarrow g(k,\bar{\eta})=0,
\label{tpcond}
\end{equation}
defining the momentum-dependent turning points
$\bar{\eta}=\bar{\eta}(k)$ is satisfied exactly, even though the
conformal time is known only at discrete points. This can be done by
locking the momentum discretization to the time discretization, i.e.,
by guaranteeing that if the time discretization is given, $k$
discretization points are chosen only if they satisfy
Eqn.~(\ref{tpcond}). Of course a predefined momentum discretization is
unnecessary if we only wish to calculate power spectra and spectral
indices in the uniform approximation; the predefined momentum
discretization is used only for initializing the mode functions in the
mode-by-mode approach, where we need the integrals on the left of the
turning point.

\subsection{Spectral Indices in the Uniform Approximation}
\label{specindapp}

The integral for the spectral index has a square root singularity at
$\eta=\bar{\eta}$ and is handled specially in the numerical
routine. We split the integral appearing in the spectral index into
two parts: 
\begin{eqnarray}
\int_{\bar{\eta}}^\eta\frac{d\eta'}{\sqrt{g(k,\eta')}}
&=&\int_{\bar{\eta}}^{\bar{\eta}+\Delta \eta}\!\!\!
\frac{d\eta'}{\sqrt{g(k,\eta')}}+\int_{\bar{\eta}+\Delta \eta}^\eta
\frac{d\eta'}{\sqrt{g(k,\eta')}},\nonumber \\ 
\end{eqnarray}
where $\Delta \eta$ is a small quantity. Note that $\Delta \eta$ is
really $k$ dependent, because the discretization in $\eta$ is not
equidistant.
 
In the first integral we can substitute $\nu^2(\eta)$ by
$\bar{\nu}^2(\bar{\eta})$, i.e., insert the leading order of the local
approximation. The first integral can then be calculated analytically
and keeps track of the inverse square root singularity, while the
second integral has no singularity and can be easily calculated
numerically.  The quantity $\Delta \eta$ is given by the actual time
discretization in physical time $t$ that we have chosen.  It is
further required that $-2\bar{\eta}>\Delta \eta$, i.e., $\Delta \eta$
be sufficiently small. As $\bar{\eta}(k)\to 0^-$ in the limit $k\to
\infty$, this relation also constrains the highest reliable mode for a
given time discretization in the exact numerical results. The first
integral gives
\begin{eqnarray}
\int_{\bar{\eta}}^{\bar{\eta}+\Delta \eta}\frac{d\eta'}
{\sqrt{g(k,\eta')}}
&\simeq& \int_{\bar{\eta}}^{\bar{\eta}+\Delta \eta}\frac{d\eta'}
{\sqrt{\frac{\bar{\nu}^2}{\eta'^2}-k^2}}\nonumber\\
&=&\frac{1}{k}\sqrt{-2\bar{\eta}\,
\Delta \eta -\Delta \eta^2}\label{eq:firstintervall}.
\end{eqnarray}

In order to avoid calculating the integrals numerically up to
$\eta\to 0^-$, we calculate the remainder of the integral from an
asymptotic value $\eta_a$, where the integrand is sufficiently small
and we can stop the numerical integration, to $\eta=0^-$, assuming
that $\nu^2(\eta)\simeq \nu^2(\eta_\mathrm{a})
+2\nu(\eta_\mathrm{a})\nu'(\eta_\mathrm{a})(\eta-\eta_\mathrm{a})$.
Then we have 
\begin{eqnarray}
&&-2k^2\lim_{k\eta\to 0^-}\int_{\eta_\mathrm{a}}^\eta 
\frac{d\eta'}{\sqrt{\frac{\nu^2(\eta)}{{\eta'}^2}-k^2}}\nonumber \\
&&\simeq -2\sqrt{\nu^2(\eta_\mathrm{a})
-2\nu(\eta_\mathrm{a})\nu'(\eta_\mathrm{a})\eta_\mathrm{a}}
+2\sqrt{\nu^2(\eta_\mathrm{a})-k^2\eta^2_\mathrm{a}}\nonumber \\
&&\quad-\frac{2\nu(\eta_\mathrm{a})\nu'(\eta_\mathrm{a})}{k}
\left[\mathrm{arcsin}\frac{2\nu(\eta_\mathrm{a})
\nu'(\eta_\mathrm{a})}{\sqrt{\Delta}}\right.
\nonumber \\
&&\hspace{2.5cm}\left. 
-\mathrm{arcsin}
\frac{2\nu(\eta_\mathrm{a})\nu'(\eta_\mathrm{a})
-2k^2\eta_\mathrm{a}}{\sqrt{\Delta}}\right],
\end{eqnarray}
with 
\begin{equation}
\Delta=[2\nu(\eta_\mathrm{a})\nu'(\eta_\mathrm{a})]^2
+4k^2[\nu^2(\eta_\mathrm{a})
-2\nu(\eta_\mathrm{a})\nu'(\eta_\mathrm{a})\eta_\mathrm{a}].
\end{equation}

Alternatively, it is possible to convert the evaluation of the
spectral index into the problem of solving a differential equation,
rather than evaluating an integral. The limit $k\eta\to 0^-$ in
Eqn.~(\ref{nsint1}) is interchangeable with a conformal time
derivative, so that the physical time derivative of the spectral index
reads
\begin{equation}
\dot{n}_S[k,\eta(t)]=-2k^2\frac{1}{a(t)\sqrt{g_S(\eta,k)}}.
\end{equation}
In order to avoid the square root singularity the integration starts at 
$\eta=\bar{\eta}+\Delta \eta$, so that the ``initial'' condition
\begin{equation}
n_S(k,\bar{\eta}+\Delta \eta)
=4-2k\sqrt{-2\bar{\eta}\,\Delta \eta -\Delta \eta^2} 
\end{equation}
includes the integral in Eqn.~(\ref{eq:firstintervall}).  It is
understood that the limit $k\eta\to 0^-$ is taken when calculating
$n_S(k)$.  The integration of the differential equation with a high
order integrator is more precise than a standard trapezoidal
rule. In contrast, as the discretization in $\eta$ is not equidistant,
higher-order integration schemes would be somewhat more complicated to
implement. However, we have verified that a standard trapezoidal
integration rule already gives sufficiently precise answers. In fact,
the local approximation for the spectral index at leading order is
quite close to the numerical nonlocal integral in the cases where the
derivative expansion is valid. 

\subsection{Conversion to Physical Units}
\label{units}
For completeness, we explain here how units are handled in
the numerical implementation.  As always, it is convenient to work in
dimensionless units, i.e., by choosing $\hbar=c=1$.  In addition, we
set the factor $8\pi G$ in the Friedmann equation in the numerical
code to unity.  These choices lead to values for the input parameters,
e.g., initial conditions and coupling constants, of order unity.  This
helps prevent numerical problems arising from very large or very
small numbers.  In order to reconvert the dimensionless units to
physical units the Hubble parameter $H$ has to be rescaled via
\begin{equation}
H_\mathrm{phys}= \sqrt{8\pi} H_0 H,
\end{equation}
where $H$ is the dimensionless Hubble parameter used in the code and
$H_0=100h\ \mathrm{km \,s}^{-1}\,\mathrm{Mpc}^{-1}$. The rescaled
momentum $k$ in physical units $h\,\mathrm{Mpc}^{-1}$ is therefore
given by  
\begin{equation}
k_\mathrm{phys}=\sqrt{8\pi}\frac{h}{100 c a(0)}k\ 
\mathrm{km \,s}^{-1}\,\mathrm{Mpc}^{-1},
\end{equation} 
with $c=2.99792458\times 10^{5}\,\mathrm{km\,s^{-1}}$.  The initial
expansion rate is $a(0)$.  Throughout the paper we have dropped the
suffix ``phys'' implying that all results are given in physical units.

Next we discuss the normalization of the amplitude of the power
spectrum.  The power spectra for scalar and tensor perturbations as
defined in Eqns.~(\ref{PS}) and (\ref{PT}) are dimensionless and
therefore not sensitive to the units of $k$.  Their amplitude is
determined fully by the parameters chosen in the inflaton potential
$V(\phi)$.  Since parameters such as the inflaton mass $m^2$ in a
chaotic $m^2\phi^2$-model are generally not known, we present the
results for the power spectra with respect to the WMAP normalization
where the amplitude of scalar perturbations is given by $|\Delta
R^2|=2.95\time 10^{-9}A$ with $A=0.9\pm 0.1$ (at $k_*=0.05/$Mpc), see
Refs.~\cite{verde,spergel}.  Using the fact that for a fixed number of
e-folds, counted from the horizon crossing of $k_*$, the parameters in
the monomial potentials simply lead to a global normalization factor
\cite{ivrevs2} we can avoid having very small numbers in the numerical
calculations and just normalize the spectra afterwards. When stating
results we give in each case the parameters in the potential
corresponding to the WMAP normalization.

\end{appendix}

\end{document}